\documentclass[onecolumn]{emulateapj}

\usepackage{overpic}
\usepackage{rotating}
\usepackage[title]{appendix}

\shorttitle{Washington Multiple Populations in NGC 1851}
\shortauthors{Cummings et al.}

\begin{document}

\title{Uncovering Multiple Populations with Washington Photometry: \\
I. The Globular Cluster NGC 1851}


\author{Jeffrey D. Cummings\altaffilmark{1}, D. Geisler, and S. Villanova}
\affil{Departamento de Astronom\'ia, Casilla 160-C, Universidad de Concepci\'on, Chile;}

\and

\author{G. Carraro}
\affil{ESO, Alonso de Cordova 3107, 19001, Santiago de Chile, Chile}


\altaffiltext{1}{Center for Astrophysical Sciences, Johns Hopkins University,
Baltimore, MD 21218, USA}


\begin{abstract}
The analysis of multiple populations (MPs) in globular clusters (GCs) has become a forefront area of 
research in astronomy.  Multiple red giant branches (RGBs), subgiant branches (SGBs), and even main 
sequences (MSs) have now been observed photometrically in many GCs, while broad abundance distributions 
of certain elements have been detected spectroscopically in most, if not all, GCs.  UV photometry has 
been crucial in discovering and analyzing these MPs, but the Johnson U and the Stromgren and Sloan u 
filters that have generally been used are relatively inefficient and very sensitive to reddening and 
atmospheric extinction.  In contrast, the Washington C filter is much broader and redder than these 
competing UV filters, making it far more efficient at detecting MPs and much less sensitive to reddening 
and extinction.  Here we investigate the use of the Washington system to uncover MPs using only
a 1-meter telescope.  Our analysis of 
the well-studied GC NGC 1851 finds that the C filter is both very efficient and effective at detecting 
its previously discovered MPs in the RGB and SGB.  Remarkably, we have also detected an intrinsically 
broad MS best characterized by two distinct but heavily overlapping populations that cannot be explained 
by binaries, field stars, or photometric errors.  The MS distribution is in very good agreement with that 
seen on the RGB, with $\sim$30\% of the stars belonging to the second population.  There is also evidence 
for two sequences in the red horizontal branch, but this appears to be unrelated to the MPs in this cluster.  
Neither of these latter phenomena have been observed 
previously in this cluster.  The redder MS stars are also more centrally concentrated than the blue MS.  
This is the first time MPs in a MS have been discovered from the ground, and using only a 1-meter telescope. 
The Washington system thus proves to be a very powerful tool for investigating MPs, and holds particular 
promise for extragalactic objects where photons are limited.

\end{abstract}

\section{Introduction:}

Globular clusters (GCs) have historically been considered quintessential Simple Stellar
Populations, but a rapidly growing body of both spectroscopic
and photometric work has shown that most, if not all, GCs are composed of multiple populations 
(MPs), with distinct chemical compositions and ages.  For decades, low-resolution 
spectroscopy of limited samples in GCs has shown evidence for CN and 
CH abundance variations within some GCs (e.g., Hesser et al. 1976; Hesser et al. 1982), but establishing 
that these were MPs and not simply effects of stellar evolution required much more detailed evidence.  MPs have 
now been observed photometrically in a large number 
of GCs including Omega Cen (Bedin et al. 2004), NGC 2808 (Piotto et al. 2007), NGC 1851 (Milone et al. 
2008, hereafter M08; Han et al. 2009, hereafter H09; and Lee et al. 2009, hereafter L09), NGC 104, NGC 
362, NGC 5286, NGC 6388, NGC 6656, NGC 6715 and NGC 7089  (Piotto et al. 2012), and M4 (Marino et al. 
2008).  In these clusters multiple red giant branches (RBGs), sub giant branches (SGBs), or even
main sequences (MSs) have been detected.  Additionally, using 8-meter class telescopes with 
multi-object spectrographs, studies of large samples of GC stars at high resolution and
signal to noise are now common and show that GCs have
intrinsically broad abundance ranges in many elements (e.g., Carretta et al. 2011a for NGC 1851).  
The Na:O anticorrelation is the best studied characteristic of GC abundance spreads, and
it has been found in virtually all clusters that have been observed with high-quality spectra of 
large samples of stars (Carretta et al. 2010).  However, recent observations of Ruprecht 106 
(Villanova et al. 2013), traditionally regarded as a GC due to its mass, age, metallicity, etc., 
found that it is the first such object that does NOT exhibit an anticorrelation or even an abundance 
spread in Na or O, or indeed any other element, although the sample size is small and more stars are 
needed to consolidate this result. 

The interplay between good photometric and spectroscopic data has been crucial in uncovering 
MPs. Clearly, spectroscopic data are superior when there exists a large enough sample of stars observed at high
resolution and signal to noise, enabling the determination of detailed abundances for a 
variety of elements. However, this technique is large-telescope-time intensive.  Photometry enjoys a hefty 
advantage in this respect, allowing much larger samples covering a wider range of evolutionary stages. 
Furthermore, it provides the atmospheric parameters required for follow up determination of spectroscopic 
abundances.  In order to properly identify MPs photometrically, the choice of filters is paramount.
To date, the key has been using a combination of filters that 
include a UV bandpass. Several studies, especially Sbordone et al. (2011) and Carretta et al. (2011b), 
have now shown that realistic abundance differences in the CNO elements, as expected between MPs, 
greatly affect UV filter bandpasses because of the strong CN, CH, and NH molecular bands present 
(Figure 1).  In contrast, the redder optical filters are 
much less affected.  Because these elements and their variations are representative of MPs, a UV filter is 
now considered crucial in order to detect MPs. The Johnson U, Stromgren u, or Sloan u filters have been used for all 
previous ground-based detections of MPs with UV photometry, but all of these filters suffer serious 
limitations. They are notoriously inefficient and are confined to a wavelength range that is very sensitive to 
both interstellar and atmospheric extinction.  Therefore, they require a significant amount of large telescope time to observe 
most GCs well, giving up a large part of the efficiency advantage that photometry enjoys over spectroscopy. 

The Washington photometric system (Canterna 1976) was designed (Wallerstein \& Helfer 1966) to
derive an accurate photometric temperature and metallicity for late-type giant stars using broad 
band filters. The $T_1$ and $T_2$ (Temperature) filters are very similar to $(RI)_{KC}$, providing 
a temperature index given by the $T_1-T_2$ color which is very similar to $R-I$. The M (Metallicity) 
filter, centered near 5000 \AA, combined with an appropriate comparison bandpass such as $T_1$, 
yields a metallicity indicator. Finally, at the time the system was being setup, the phenomenon of 
CN and CH variations within GCs was being discovered (e.g. Hesser et al. 1976) and it was realized 
that the addition of another filter designed to measure the CN/CH strength independently of the 
metallicity would be of great use in disentangling these two abundances. \textit{Thus, the C (Carbon) filter 
was introduced, specifically to search for MPs.} To our knowledge, this was the first and still only 
such filter so designed. In order to maintain the broadband characteristics of the system, the filter 
needed to cover a wide wavelength range that included significant CN and CH bands. A design goal was 
to include both the UV CN band near 3600 \AA\, as well as the CH G-band near 4300 \AA. As shown in 
Figure 1, the Washington C filter covers the same bandpasses and molecular bands as the other UV filters 
used to uncover MPs, and in fact more of these latter than any other competitor.  It is also 
both much broader (FWHM $>$ 1000 \AA) and centered significantly redward ($\sim$3900 \AA) of the 
other filters (see Table 1), making it much more efficient and less reddening and extinction sensitive.  It even 
includes some sensitivity to the NH band at 3360 \AA. Note that C has a very similar blue response 
as Johnson U but is centered 300 \AA\, redder and retains significant sensitivity to beyond 4500 \AA, 
far beyond the red cutoff of U. The peak transmission is also higher in C than the other filters.
It is thus much faster (typically $>$3 times faster at obtaining the 
same S/N) than U and is many times more efficient than Stromgren or Sloan u, especially in red giants.

\begin{center}
\begin{longtable}[htp]{c c c c c}
Filter & Central $\lambda$ & FWHM ($\lambda$) & Peak Transmission & Source\\
\hline
Johnson U & 3570 & 650 & 72.47\% & 1 \\
Washington C & 3850 & 1075 & 83\% & 1 \\
SDSS u & 3600 & 400 & 65.49\% & 1 \\
Stromgren u & 3537 & 278 & 38\% & 2\\
\hline
\caption{The Washington C filter is shown to be centered redward and significantly broader than the
other UV filters.  These are filter specifications taken from (1) the CTIO MOSAIC Filter Set and from (2) 
The KPNO Filter set.  Minor variations of all these parameters occur between different copies of the 
same filter.}
\end{longtable}
\end{center}

The Washington system has been used for a wide variety of studies. Originally (Canterna 1976), metallicity 
determinations were based on a mean of the abundances derived from both the M and C filters as long as 
they were in good agreement. If there was a significant difference, the M metallicity was used and the star 
was labeled as having a CN/CH excess. This led to the discovery of a number of such stars, including the 
first extragalactic example (Canterna and Schommer 1978). Subsequently, Geisler (1986) discovered that 
the metallicity sensitivity of the C filter was superior to that of the M filter and recommended that 
future studies preferentially utilize the C filter for measuring metallicity.  Indeed, this filter is sensitive 
to both CN/CH as well as traditional metal abundance because the bandpass also includes such features as the 
Ca II H and K lines and a series of strong Fe I lines spanning from 3400 to 4400 \AA.  Thus, the ability of 
the system to search for MPs was never fully exploited, despite its clear advantages in terms of efficiency
and specific design for uncovering MPs. 

Recently, a C-filter equivalent, F390W, has been added to the WFC3 filter complement aboard HST.
This filter has indeed been utilized in MP studies, e.g. Bellini et al. 2013, and proven quite successful
in this regard. However, such studies have not recognized or treated this as the Washington C filter.  
To our knowledge, the only relevant study to test this filter as the Washington C filter is a very 
recent one by Ross et al. (2014), where they tested the ability of various HST/WFC3 filters to derive 
photometric metallicities in five well studied galactic star clusters.  They found the F390W 
filter is both the most efficient and also provides the second-best metallicity index after the F390M 
filter, which is unfortunately 4.5 times narrower. This study proves that the C filter is indeed very metallicity 
sensitive but does not demonstrate its potential for the MP phenomenon.  Likewise, there have been no
ground-based studies with the Washington system investigating its utility in this regard.

Given the above, we felt an initial ground-based study to investigate the ability of the Washington 
system to uncover MPs in Galactic GCs was well-motivated. Our initial target is NGC 1851.  NGC 
1851 is a particularly interesting cluster in many respects. Photometrically, it displays two RGBs 
(H09) and SGBs (M08; H09), and it has both a red horizontal branch (RHB) and a blue horizontal 
branch (BHB), which is not commonly seen in GCs.  A great advantage for studying this cluster is 
its very low reddening of E(B-V) = 0.02 (Harris 1996), meaning that the potential complicating 
effects of variable reddening can be ignored (NB: $(E(C-T_1)\sim 2E(B-V)$).  Spectroscopically, 
the two RGBs observed in NGC 1851 typically exhibit different abundances in a variety of elements, 
most strikingly in Na, Ba, and N (Villanova et al. 2010, hereafter V10; Carretta et al. 2011a, 
hereafter Ca11; Carretta et al. 2014, hereafter Ca14), and the two SGBs appear to correspond 
to high and low Ba abundances (Gratton et al. 2012a).  Therefore, abundance differences likely play 
an important if not dominant role in creating the separate sequences.  The details for several
of these key abundance differences are still debated.  For example, V10 find that in 15
RGB stars the total C+N+O content shows no significant variation, while in a more limited sample of 
4 giants Yong et al.
(2009) find that the C+N+O content may vary by up to a factor of 4.  Additionally, it has been 
argued that these two populations show evidence of different radial distributions by Zoccali et al. 
(2009) and Ca11, but both Milone et al. (2009) and Olszewski et al. (2009) have argued that there 
are still two sequences in the SGB at large radii with no statistically significant variations in 
the number ratios of the two populations with increasing radii.  A further intriguing property of 
NGC 1851 is that it is one of the few GCs that possesses an apparent real spread in the heavy 
elements (e.g., Fe), not just in the light elements (Ca11).  Theories to explain the MPs in NGC 1851 
include: 1) an initial population formed and ejecta from its high-mass stars, including SNe, polluted 
the remaining gas.  Soon after ($<$1 Gyr), a second population formed from the polluted gas that was 
not expelled from the cluster (see M08, Ventura et al. 2009, and Joo \& Lee 2013).  2) There was 
a merger of two GCs of slightly different age and composition (see Ca11).  

This paper is the first in our survey of a sample of southern GCs using the Washington filters to 
investigate MPs. It is organized as follows:  In Section 2 we discuss our observed data and methods 
of analysis.  In Section 3 we discuss our photometric results and analyze in general what can be 
accomplished with them for studying multiple stellar populations in GCs using relatively little 
telescope time.  In Section 4 we search for multiple MSs using our photometry.  In 
section 5 we discuss in more detail the structure we observed in the red HB.  In Section 6 we 
analyze potential differences in the radial distributions of the two populations.  Lastly, in 
Section 7 we summarize our results and conclusions, and we point in several interesting directions 
for future analysis.

\section{Observations and Analysis:}

In order to demonstrate the efficiency of the Washington system in MP studies, we decided to
use a 1-meter class telescope. Our observations of NGC 1851 were performed at the SWOPE 1-meter 
telescope at Las Campanas Observatory using the SITe\#3 detector with 2048x3150 pixels at 
0.435''/pixel and a field of view of 14.9x22.8 arcminutes.  Both the R and T2 observations were 
performed during grey time on October 21, 2011 with one short C image taken that night before the 
Moon rose, and the remainder of the C observations were performed before the Moon rose on
October 25, 2011.  Note that Geisler (1996) has shown that the $R_{KC}$ filter is an 
accurate and much more efficient substitute for the Washington $T_1$ filter.
Both nights were determined to be photometric based on a series 
of observations of Geisler (1996) standard star fields.  All of the R and T2 observations
plus the single C image from October 21, 2011 had a FWHM of 0.95'' to 1.05'', while the 
remaining C observations from October 21, 2011 had a FWHM of 1.5'' to 1.95''.  A total of 
2 short C (300 seconds each), 9 long C (1200 seconds each), 1 short R (100 
seconds), 3 long R (400 seconds each), 1 short T2 (300 seconds), and 3 long T2 (1200 
seconds each) images were obtained, giving a total of 11,400 seconds for C, 1300 seconds 
for R, and 3900 seconds for I.  In comparison, the Johnson UVI observations of H09
had a total exposure time of 5239 seconds on the Blanco 4-meter telescope (versus our 1-meter).
Scaling the total exposure times by the aperture of the telescopes gives that this is
effectively $\sim$5 times more telescope time.  In comparison to the Stromgren observations 
of Lee et al. (2009), they had a total exposure time of 49,710 seconds (nearly 14 hours) 
on the SMARTS 1-meter telescope, which is $\sim$3 times more telescope time.  

Standard IRAF tasks were used to process the data.  Based on the detailed tests from 
Hamuy et al. (2006), we have adopted their non-linearity corrections for SITe\#3.  We 
also took a series of images to test for shutter corrections of the
detector, but we found that if we corrected the shutter test images for the adopted
non-linearity of the detector, no significant shutter correction was necessary.  Since
the NGC 1851 field is crowded, DAOPHOT was first applied independently in 
each image to a sample of bright and isolated stars to 
determine a first approximation for a quadratically-varying PSF.  This initial PSF was then 
applied to subtract all neighboring stars from a larger sample of PSF stars ($\sim$200 
stars), which were used to determine a final quadratically-varying PSF (Stetson 1987).  
These final PSFs were applied with ALLFRAME to self-consistently measure the photometry of 
all images simultaneously (Stetson 1994).  Aperture corrections were determined by comparing 
the PSF photometry of bright (but unsaturated) stars to their aperture photometry 
after all nearby-neighbor stars had been subtracted.  Stars spanning the entire field 
were used to test for spatial dependence of the correction, but it was found to be uniform 
across the field in all filters.  Additionally, several cuts were applied to the photometry 
based on the magnitude error, the chi-squared, and the sharpness value output by ALLFRAME.  
Individual star measurements with a magnitude error greater than 0.15, a chi-squared greater than 
2.5, or an absolute value of sharpness greater than 1 were cut from the results.  Lastly, 
the brightest stars that were affected by non-linearity of the detector, determined by 
where magnitude errors began to sharply increase with increasing brightness, were also cut.

The individual high-quality C, R and T2 magnitudes from ALLFRAME were first combined using
DAOMASTER to create final C, R, and T2 instrumental magnitudes, which were then matched and 
combined to create a final list of stars with 2 to 3 of these filter magnitudes.  The 
observations were transformed to the standard Washington system based on our observations 
of the standard star fields published in Geisler (1996), where the R magnitudes have been 
transformed to T1.  For each night we had 8 standard observations in all 3 filters from 5 
different standard fields.  This gave a total of 79 standard stars in C, 70 in T1, and 69 
in T2 that covered a broad range in UT, airmass, magnitude, and color.  The RMS of the 
standardization was 0.024 for C, 0.016 and 0.017 for T1 when using a T1-T2 color term and 
a C-T1 color term, respectively, and 0.013 for T2, indicating both nights were of excellent 
photometric quality. Both C and T2 were well behaved and 
only required linear color terms.  T1 required a quadratic color term both nights, but 
this quadratic color term has consistently been found across multiple nights of photometric 
observations and is likely due to the difference in the width, but similar centers, of 
the T1 and R filters.

Due to the large pixel scale of the SITe\#3 camera (0.435''/pixel), we found that the 
very good seeing on October 21, 2011 produced stars with FWHMs of only $\sim$2.1 pixels.  
This created minor undersampling issues with our PSF measurements that increased the errors 
of the R and T2 observations, which were all obtained on that night.  Using in DAOPHOT and 
ALLFRAME a fitting radius 0.4 pixels smaller than the measured FWHM was shown to moderately 
improve the errors, so we have used those measurements to produce our R and T2 instrumental 
magnitudes.  Conversely, the C observations that were predominantly performed on October 
25 had FWHMs of $\sim$3 to 4 pixels, so the C photometry was not affected by the large pixels.

Spatial variations of the photometric zero-point are a common problem that can be caused, for
example, by focus variations across the field, and to derive higher precision photometry we
have corrected them.  By generating a variety of color magnitude diagrams using our C, T1, and T2 
photometry and the B and V photometry acquired through private communication with Momany Y. (observed at 
the 2.2m ESO with WFI camera), we have used the RGB and the upper MS to fit fiducials and test for spatial 
variations in the relative color distributions.  This method is similar to that typically applied for
correcting differential reddening, but we do not apply a reddening law.  We can assume that all 
color variations are due to true photometric variations and not variable reddening because NGC 1851 has a
very small reddening of only 0.02 (see Section 1).  Comparisons between our own C, T1, and T2 
photometry show that there are meaningful spatial variations with respect to each other while the 
comparisons of B and V show no significant spatial variations with respect to each other.  Therefore, 
to determine the variations in our magnitudes we have used V as the reference magnitude by creating 
colors with our filters versus their V and have only used their V for our magnitude axis.  We can 
assume that all resulting spatial variations in color are due to our photometry and correct it by 
fitting the smooth spatial variations and apply this correction directly to our magnitudes.  

The magnitude of the final corrections in each filter are typically
or order 0.015 in C and 0.02 in both T1 and T2 and symmetric around zero.
The corrections are comparable in magnitude across a majority of the field, but it should be noted that 
near only the northern edge of the image the corrections quickly become larger in all filters,
being of order 0.06 in C and 0.05 in both T1 and T2.  This variation at the northern edge is not
of major concern considering the detector spans nearly 23 arcminutes along this axis and members of 
the centrally placed NGC 1851 will be very sparse near this edge.
The independently corrected C and T1 magnitudes are both self-consistent with each 
other and show consistency in all magnitude ranges across the full field of view.  However, the 
T2 magnitudes show spatial variations that are moderately inconsistent between the brighter and 
the fainter stars.  We have chosen to correct these spatial variations differently for the brighter 
(T2$<$18) and fainter stars (T2$\geq$18) in T2.  The source for these inconsistencies at differing 
magnitudes is uncertain, but it is possibly related to the minor fringing that occurs only in the 
T2 images.  We did not have appropriate reference images to fully correct this minor fringing, but our 
spatial corrections will have also corrected its largest effects while the small scale fluctuations 
will have only introduced minor but not insignificant additional T2 errors.  With this greater 
photometric uncertainty, we will still present our T2 photometry in this paper, but will also 
supplement our photometry with the B and V photometry of Momany Y. (private communication) to 
provide additional non-UV magnitudes.  As a further constraint, we will not base any conclusions 
on characteristics seen only in color-magnitude diagrams (CMDs) 
containing T2, and will show that all characteristics observed using the T2 magnitudes are 
qualitatively consistent with the other comparable filter combinations.

For our final photometry, see Table~2 for an example, we have also applied a strict cut of our data near the core because both the spatial 
variations of the errors and our measurement tests of artificial stars show that crowding begins to 
affect an increasingly large fraction of the stars that are within 2 arcminutes of the center.  The effects of
crowding are not reliably represented by the ALLFRAME determined photometric error but are better represented
by the resulting $\sigma$ of the multiple measurements of each star.  Therefore, each star near the center has
been cut unless it has been measured multiple times in each filter and the resulting $\sigma$ of the multiple 
measurements is $<$0.02 in both filters.  This removes stars affected by crowding without removing the 
bright and centrally concentrated RGB stars that are still well measured near the center.  Additionally, 
because we only have one available short T1 image, no stars brighter than a T1 of 15 will have multiple T1 
detections.  Therefore, for these stars we only require that they have multiple C detections and a resulting 
$\sigma$ of less than 0.02.  This effect from crowding in the core prevents us from having
any overlap in our final sample and the proper motion sample of M08, where they looked at the SGB and MS in 
the central 2.7 arcminutes.  However, their analysis does demonstrate that for NGC 1851 the field-star 
contamination is quite minor, as expected from its galactic latitude of -35 degrees.

\vspace{1cm}
\begin{sidewaystable}
\begin{center}
\begin{longtable}[htp]{c c c c c c c c c c c c c c c}
ID & X & Y & C & C Err & C Disp & T1 & T1 Err & T1 Disp & T2 & T2 Err & T2 Disp & C Count & T1 Count & T2 Count\\
\hline
16798	& 1330.59	& 1083.73	& 15.971	& 0.005	& 0.004	& 12.833	& 0.016	& 0   	& -	& -	& -	& 4 	& 1	& 0\\
16928	& 1274.17	& 1675.08	& 16.861	& 0.005	& 0.008	& 15.738	& 0.015	& 0.006	& -	& -	& -	& 4 	& 3	& 0\\
6984	& 633.17	& 1725  	& 17.868	& 0.004	& 0.014	& 16.304	& 0.007	& 0.007	& 15.811	& 0.012	& 0.013	& 11	& 4	& 4\\
2832	& 1295.54	& 1435.55	& 18.55 	& 0.004	& 0.017	& 17.228	& 0.01	& 0.009	& 16.728	& 0.014	& 0.012	& 10	& 4	& 4\\
6953	& 1349.92	& 1721.71	& 18.757	& 0.005	& 0.013	& 17.831	& 0.017	& 0.008	& 17.41 	& 0.023	& 0   	& 11	& 4	& 2\\
6345	& 1194.57	& 1673.08	& 19.48 	& 0.007	& 0.019	& 18.469	& 0.022	& 0.044	& 18.087	& 0.025	& 0.014	& 11	& 3	& 3\\
3760	& 1622.07	& 1513.54	& 19.657	& 0.004	& 0.015	& 18.664	& 0.009	& 0.002	& 18.26 	& 0.012	& 0.006	& 11	& 4	& 4\\
\hline
\caption{The Washington Photometric Catalog of NGC 1851.The columns are the stellar ID, XY positions, and the C, T1, and T2 final magnitudes.  For each
magnitude both the photometric error and the photometric dispersion are given.  The final 3 columns give 
the total number of independent observations performed for that star in each filter. Table 2 is published in its entirety in the electronic 
edition of the {\it Astronomical Journal}.  A portion is shown here 
for guidance regarding its form and content.
}
\end{longtable}
\end{center}
\end{sidewaystable}

\vspace{1cm}
\section{Multiple Populations in the Red Giant Branch and Subgiant Branch:}

Figure 2 shows 4 CMDs from our final photometry.  We show representative error bars on the side 
placed at each magnitude.  
These errors have been determined by both types of errors output by DAOMASTER: the combined photometric measurement 
error output by ALLFRAME and the $\sigma$ based directly on the observational scatter across the multiple images.  
We typically find the photometric error dominates in the brightest stars but the observational scatter
dominates in the fainter stars, and for each magnitude we take the largest of these two errors to be the
better representation of its final error.  For errors in color we add in quadrature the final errors from each 
input magnitude.  Lastly, the representative errors shown in Figure 2 are found by taking the median error of 
all stars in a 1 magnitude range.  

The brightest giants are saturated in our observations, so we have expanded our sample by taking 
advantage of the short photometric observations of NGC 1851 in C and T1 by Geisler \& Sarajedini 
(1999) and the I observation from Momany Y. (private communication).  Detailed comparisons of our 
observations to those of Geisler \& Sarajedini (1999) show that their magnitudes are systematically 
brighter than ours by 0.006 in C and 0.029 in T1.  Therefore, we have offset their results to be 
consistent and have used x symbols in the lower-right and upper-right panels of Figure 2 to 
distinguish their data.  For expanding our T2 magnitudes, we have used the brightest $\sim$1000 stars 
(14.4$<$T2$<$18.5) that have both our T2 and I from Momony to create a transformation relation.  Because 
of the strong similarities between these two filters the relation is linear across this full magnitude
range with little scatter, which allows us to extrapolate to stars brighter than T2=14.4 without likely 
introducing significant systematics.  Using this relation we transform the I magnitudes for the brightest
stars (T2$<$15) not in our sample to T2, and we mark this data in red in the upper-left 
and lower-left panels of Figure 2.

In the T1-T2 CMD (upper-left panel of Figure 2) there are no clear, separate branches visible, and the 
widths of the RGB and MS are similar to the representative errors displayed.  This is consistent with 
previous ground-based studies that have shown that colors without a UV filter do not show either multiple 
branches or even a significantly broadened RGB.  In contrast, things are very different in all CMDs including 
the C filter.  We plot the C-T1 color both against T1 (upper right) and C (lower right).  
In both CMDs we see that in addition to the primary population there is a smaller but still significant 
population of SGB stars that are fainter than the primary SGB and a population of RGB stars that are redder 
than the primary RGB.  The SGBs are shown more clearly in the insets.  In Figure 3 we focus on the RGB and SGB
in C-T1 versus C and for clarity color the two RGB branches.  Both the fainter SGB and the redder RGB 
appear to be related and to create a continuous branch, a second population.  These observations 
of a second population are similar to those of H09 using Johnson U and I.  In direct comparison to H09,
it should be noted that our second population stars do not appear to create as distinct or as well  
defined of a sequence as found in H09.  They also appear to be less well defined than our primary
population stars, suggesting that this is not the result of our observations having larger errors.  Therefore, 
we cannot argue that in C the second population creates a photometrically distinct sequence from the primary 
SGB and RGB.  Similar to C, however,
the Stromgren observations in L09 also do not show a distinct second population sequence in the RGB.

In further comparison to the U-I versus U observations of H09, the mean color separation of the 
red and blue RGBs are quite similar to ours.  Above the observed RGB bump, where the separation is the 
largest, both their observations and ours in C-T1 versus C show color separations of $\sim$0.25.  
In the lower RGB, while our color separation is more difficult to define here, we both find a color
separation of $\sim$0.15.  This significant 
change in color separation between the two populations between the upper and lower RGB is of interest.  
In our observations this change is more appropriately analyzed in C-T1 versus T1 (upper-right panel of 
Figure 2) because here the two populations are shifted in color but not magnitude.  We see in C-T1 versus 
T1, in comparison to C-T1 versus C, that the true change in color separation between the upper and lower RGB is 
smaller but still is of significance.  We expect a color separation increase, however,
because the brighter giants are also increasingly cooler, leading to stronger molecular bands, giving that 
identical CNO variations will produce greater C magnitude differences.

Comparing the numbers of our red and blue RGB stars marked in Figure 3 gives that 
the number of the clearly distinct red RGB stars is only $\sim$13.6\% of the total RGB population.  This is 
significantly smaller than the $\sim$30\% typically found for the second population when performing observations 
using Johnson filters (M08; H09), where the second populations in the SGB and RGB appear more distinctly separated.
Therefore, this may be related to the more diffuse nature of this second population in our observations.
This discrepancy and the second population's color distribution will be discussed in more detail in Section 4.2.  
For simplicity we will refer to these distributions from now on as the red and the 
blue RGB branches, and the bright and the faint SGB branches.  

In the C-T2 CMD we can still clearly see the separate branches of both the SGB and the RGB, but it is of 
interest that in this color we can no longer clearly detect the separate RHB branches 
observed with C-T1.  Additionally, the two SGB and RGB branches are less defined.  These differences
may more be related to the increased errors for T2, rather than differing
characteristics between T1 and T2, but these factors and the significantly shorter exposure times
required for the T1 filter gives a clear advantage for the T1 over the T2 filter.  Therefore, C-T1 
will be our primary color for our analysis in the rest of this paper.  
But it remains ideal to still observe both T1 and T2 to verify the observed populations with 
both filters versus C and to compare these colors to the narrow RGB that is shown only in T1-T2.  

We have further analyzed the significance of these photometric features in the RGB by comparing the color 
errors of the stars to their color differences from the RGB fiducial.  We do this in 1 magnitude 
bins based on the median values of these parameters.  Table 3 shows our results.  In T1-T2, the 
ratio of width to errors ranges from 0.83 to 1.18, so the color errors are comparable to the 
observed width of the RGB  and further strengthens our finding that there is no meaningful 
broadening in T1-T2.  We perform a similar comparison for C-T1 using C-T1 versus T1.  Across the 
full range of magnitudes the RGB appears to be intrinsically broad.
In all 5 magnitude bins the width to error ratio is meaningfully larger than that seen in T1-T2, ranging 
from 1.5 to almost 3 $\sigma$.  The comparison for C-T2 versus T2 also shows that on the 
RGB the width is significantly larger than the error, at 1.3 to 2.6 $\sigma$.  The width of the
broadened RGBs is the least significant in the faintest (hottest) RGB stars where the molecular bands
will be the weakest.  However, even at their weakest the RGB in C-T1 or C-T2 is still 
broader than that observed for T1-T2.  

\begin{center}
\begin{longtable}[htp]{c c c c}
Magnitude Range & Median Width & Median Error & Width to Error Ratio \\
\hline
\multicolumn{4}{c}{RGB in T1-T2 versus T1}\\
\hline
15-16 & 0.018 & 0.022 & 0.83\\
16-17 & 0.023 & 0.020 & 1.18\\
17-18 & 0.019 & 0.019 & 1.04\\
\hline
\multicolumn{4}{c}{RGB in C-T1 versus T1}\\
\hline
13-14 & 0.035 & 0.023 & 1.50\\	          
14-15 & 0.042 & 0.016 & 2.69\\	
15-16 & 0.039 & 0.014 & 2.77\\
16-17 & 0.032 & 0.017 & 1.95\\	
17-18 & 0.028 & 0.018 & 1.55\\          
\hline
\multicolumn{4}{c}{RGB in C-T2 versus T2}\\
\hline
14.4-15 & 0.056 & 0.022 & 2.59\\
15-16 & 0.035 & 0.021 & 1.64\\
16-17 & 0.034 & 0.020 & 1.70\\
17-17.9 & 0.027 & 0.021 & 1.34\\
\hline
\caption{Comparison of RGB widths and errors in the available colors at specified
magnitude ranges.}
\end{longtable}
\end{center}

Figure 4 shows 4 additional CMDs based on combinations of our photometry and the B and V photometry.
In the upper-left panel we look at V-T1 versus T1, which as expected is very similar to the T1-T2 
versus T1 CMD from Figure 2.  The more limited errors given for the B and V magnitudes do not allow us to 
create a detailed and consistent presentation of representative errors, but similar to what is seen
in T1-T2 the RGB does not present multiple branches or appear meaningfully broadened in this color.  In the
lower-right panel we look at C-V versus C, and this CMD shares many similarities with both the
C-T1 and C-T2 versus C CMDs from Figure 2.  In particular, the RGB is heavily broadened with a 
more populous blue RGB and a sparser red RGB, and a faint SGB branch is also seen.  The upper-right
panel shows the very interesting B-T1 versus T1 figure.  While the color does not involve the key C 
filter, the Johnson B filter overlaps with the C filter and still contains some of the important CN 
and CH bands (see Figure 1) that are believed to create the C magnitude variations.  Consistent with 
this, the B-T1 versus T1 CMD exhibits a less significant but detectable population of red RGB stars 
that are similar to that seen in C-T1 versus T1.  This leads us to the lower-left panel, which 
analyzes the C and B magnitudes using C-B versus C.  Consistent with the previous observations,
the RGB still is broadened, but less so than that observed in C-V.  This is because both the C and B 
magnitudes are affected by the CNO variations, leading to a weaker but still detectable color difference
between the two populations.

\section{Multiple Populations on the Main Sequence:}

\subsection{Outer Annulus Analysis:}

We can also search for MPs on the MS of NGC 1851, which have never been detected before, despite
even being searched for in HST data (see M08).  To limit our errors, we will first
concentrate on stars well outside of the dense core but not apply too strict of an error cut or we
will lose the faint MS stars.  In Figure 5 we only show stars that have a C-T1 color error of 
$<$0.05 and that are in an annulus around the cluster center from 4 to 6.5 arcminutes.  The 
half-mass radius from Harris (1996) is 0.52 arcminutes, giving that this is 7.7 to 12.5 half-mass radii 
from the center.  Additionally, the representative error bars shown have been updated based on 
this subsample.  We have plotted all of the colors versus T1 to increase the color difference between two potential
populations.  If we plotted versus C, a second redder population would also be shifted fainter, partly
parallel with the MS, and decrease the apparent color difference at constant magnitude.  Remarkably, in 
the left panel of Figure 5 we see clear signs for a broad MS with a substantial wing extending to the 
red that is not seen to the blue.  This moderately redder MS population follows the same general shape as 
the primary population.  Additionally, at the primary population's turnoff this red population maintains 
its color separation and does not simply merge into the turnoff.  Indeed, it appears to transition 
into the faint SGB that we have previously identified.  
To help illustrate this second population we have colored its likely members red in Figure 5, 
where we have based this solely on C-T1 and for stars with T1$>$19. 

We must further analyze this red branch to test if it is a true second population or is simply a binary 
sequence, but we must also consider if it is a combination of both.  In the left panel of Figure 5 we 
first characterize the primary MS with a solid-black line.  We also illustrate a corresponding equal-mass 
binary sequence with a black-dashed line, which characterizes where the upper envelope of the 
binary sequence would be.  This shows that the binaries
would merge into the primary turnoff and then have their own turnoff at a brighter magnitude.  Comparing to
our data we find at the primary turnoff, even when considering errors, our observed second population 
still remains distinct and moderately red, and at fainter magnitudes the observed stars are increasingly 
too blue to be consistent with a binary sequence.  Furthermore, there are a negligible number of stars 
that are consistent with a second brighter turnoff of binaries.  

In Figure 6 we further illustrate the 
characteristics of a binary sequence using population synthesis of a $\it{single}$ population with a binary
fraction of 3\% using the TRILEGAL code (Girardi et al. 2005).  We have input errors in this population 
synthesis based on our observed relation 
between photometric error and magnitude.  To help distinguish the binaries in Figure 6 we have colored
red all binaries with a mass fraction $>$0.6, while binaries of lower mass fraction will not be
photometrically distinguishable from single stars.  This population synthesis in C-T1 further shows that 
the binaries will appear relatively very red at fainter magnitudes, quickly begin to converge in color at 
brighter magnitudes and merge into the primary turnoff, and then have their own weakly populated turnoff
at brighter magnitudes.  In contrast to this, a true second population would be shifted redder
in C-T1 but have no meaningful change in T1 magnitude.  The median C-T1 color difference between these two
branches is 0.096, and in Figure 5 the solid-red line demonstrates that this color shift is more comparable to
the red MS stars.  That the brighter stars are moderately bluer and the fainter stars moderately redder than
this uniform color shift is further consistency with the two populations in the RGB stars (see Figure 2), 
which also exhibited changing color separations.  In both cases the cooler stars, which will have stronger
molecular bands, will create increased color separations at equivalent CNO abundance variations.

In the center and right panels of Figure 5 we will now analyze this observed second MS branch in other 
colors by taking the selected sample from C-T1 (colored red) and also coloring them red in these CMDs.  
In C-T2 (center panel of Figure 5) we see there is still a prominent redder population that is primarily 
composed of the same red stars observed in C-T1.  The scatter of the two populations appears larger in 
this color, but this is consistent with T2's moderately greater error.  The median C-T2 color difference 
between the marked black and red populations is 0.109 (similar to the 0.096 in C-T1), and similarly shows 
the color difference between the two populations moderately increases in the fainter (cooler) stars.  
Comparing to the population synthesis for C-T2 (center panel of Figure 6), we see the same binary 
characteristics found in C-T1 and the same differences when compared to our observed second branch in C-T2.  
However, there is one key difference between the MS in C-T1 in comparison C-T2 that becomes readily apparent in 
their synthetic binary sequences: across the same T1 magnitude range the MS changes color in C-T2 more 
rapidly with decreasing magnitude, which greatly increases the apparent color separation between the single 
and binary sequences.  Therefore, the greater observed color dispersion in the red branch of C-T2 
(center panel of Figure 5) may be indicative of at least minor binary contamination, but overall 
because the color difference between the two populations in C-T1 and C-T2 are in agreement, this suggests 
that this is predominantly a true second population created by a shift in C magnitude.

We now mark these observed red-population stars in T1-T2 (right panel of Figure 5) and see that in this 
color there is no clear difference.  The red-population stars observed in both C-T1 and C-T2 are distributed 
remarkably uniformly within this single sequence, and the median T1-T2 color difference is insignificant at 
0.014.  In contrast to this, the right panel of Figure 6 shows that in T1-T2 the binaries will still remain 
meaningfully redder than the single-star sequence.  However, in our T1-T2 observations we do note a small number of color 
outliers are still seen in both populations.  The total T1-T2 color distribution has a $\sigma$ of 0.026 and
we define the outliers as stars more than 2.5$\sigma$ from the central sequence.  2.6\% of the primary population 
(black stars) are T1-T2 color outliers, with 1.1\% being blue outliers and 1.5\% being red outliers.  In comparison 
a more significant 10.4\% of the red population are outliers, but with only 2.3\% being blue outliers and a much 
larger fraction of 8.1\% being red outliers.  Therefore, based on the small number statistics, in the primary 
population there is a comparable number of 
red and blue outliers, which are also comparable with the number of blue outliers in the red population.  
This suggests that $\sim$3-4\% of both populations are true photometric outliers, non-members, 
or stars with abnormally large errors, which will approximately be distributed evenly between 
red and blue outliers.  Correcting for this there remains an additional $\sim$6-7\% of the red population 
that are red outliers in T1-T2.  This small 
fraction of the red population ($\sim$0.65\% of the total population) may be true binaries 
that have a large mass ratio and will appear to be redder than the primary single-star population in all 
colors, while 93-94\% of the red population remains strongly consistent with a second population of
differing C magnitude.  

If this second population of stars is not caused by a binary sequence, why do we not see a meaningful
population of binaries in this outer annulus of NGC 1851?  The recent definitive analysis of binary
populations in GCs by Milone et al. (2012a) may provide an answer.  They observed the MS in the central 
region ($<$2.5 arcminutes from the center) with ACS/WFC in F606W and F814W, very similar to T1 and T2, giving
a color that should make our two purported populations virtually indistinguishable in the MS.  They observed 
that NGC 1851 only has a binary fraction of 0.8$\pm$0.3\% for binaries with a mass ratio of greater than 0.5.  
Binaries with mass fractions of less than 0.5 were not able to be differentiated from single stars with their 
observations.  Based on their models they estimate that NGC 1851 has a total binary fraction of only
1.6$\pm$0.6\% in the observed central region.  This is one of the lowest binary fractions of the 59 GCs 
they analyzed.  
Furthermore, their observations find that a majority of the observed GCs, including NGC 1851, 
have decreasing binary fractions at increasing distance from their center.  This is 
consistent with expectations that binary systems, which dynamically act like a single star with mass equal 
to the sum of the binary components, will over time become more centrally 
concentrated in comparison to the less massive single stars.  All of these factors suggest that in 
our observations of this outer annulus from 4 to 6.5 arcminutes, the binaries that would meaningfully contribute 
to a binary sequence (mass ratio$>$0.5) is less than 1\% of the total population.  This is remarkably
similar to our estimate from Figure 5 that in this outer annulus only $\sim$0.65\% of our MS stars 
are consistent with large-mass-ratio binaries.  

The number ratio for the red population shown in Figure 5 is 9.5\%, which is comparable to 
the 13.6\% found for the red RGB in comparison to the blue RGB in Section 3.  Additionally, a comparable analysis 
of the SGB subsample shown in Figure 5 gives that the faint branch of the SGB is 12.7\% of the total population.  
The agreement of the number ratios in all 3 regions of the CMD further strengthens the conclusion that 
this extended red population in the MS is a second population consistent with the one observed in both the SGB 
and RGB.  However, it should be noted again that these number ratios are based on the photometrically distinct red stars, 
and more reliable color distributions and population ratios will be discussed in Section 4.2.  

To further broaden our analysis, we can use the available B and V magnitudes.  In Figure 7 we show the same sample 
of stars from Figure 5 (when B and V are available) using C-V, B-T1, and B-V.  As in Figure 5 we again have marked 
the red population from C-T1 in all 3 CMDs.  In the left panel of Figure 7 we see that in B-T1 the red MS population 
is not as distinct as in C-T1 or C-T2, but it is still significantly redder than the primary MS by 0.076.  This is 
similar to what was seen with the RGB in B-T1 (Figure 4), which illustrated that in this color the two populations 
still exhibit a significant but smaller color difference to that seen in colors with C.  As with the RGB, this is 
expected since CNO variations will also cause 
moderate differences in B magnitudes.  The C-V color in the central panel is more surprising because even
though the red MS still is consistently redder than the primary sequence, the difference is comparably weak 
at only 0.052.  Lastly, in B-V we see that while there is not a large difference in color between the two populations, 
the red MS is still moderately redder than the primary MS by 0.030.  This is relatively a minor difference but
it is not insignificant.  Both the central and right panels of Figure 7 suggest that for these two MS populations 
there may also be a weak difference in V magnitude, which results in a weaker than expected color difference when 
comparing to both C and B.  Could this apparent difference in V magnitude for these two populations be indicative of 
additional effects beyond CNO variations for the two populations, e.g., possible differences in helium abundance, 
metallicity, or a large total C+N+O difference between these two populations?  

This red MS population is very unlikely to be due to field stars, as it is centrally
concentrated (see Section 6), and the field contamination is shown to be relatively minor in
the proper motion analysis by M08.  Stars photometrically consistent with binaries appear to only 
represent a minor portion ($\sim$6-7\%) of this red MS, consistent with this cluster's small binary
population (Milone et al 2012a).  Furthermore, the effects of photometric error and crowding
do not seem to be artificially creating this second branch (see the Appendix for detailed
discussion of this).  Therefore, Figures 5 to 7 provide strong evidence that these two MS 
branches are predominantly two populations of differing C magnitude with no meaningful 
difference in either T1 or T2, qualitatively consistent with the second population observed in both the 
SGB and RGB.

\subsection{Full Field Analysis:}

With strong evidence for two MS populations found in an outer annulus subsample of $\sim$1000 main-sequence stars, 
it is of great interest to look at a larger sample covering nearly the full field.  Beginning with all stars from our original 
sample shown in Figure 2, which only placed a stringent error cut on the stars within 2 arcminutes of the center, we 
will focus on the 3162 MS stars with T1$>$19 and C-T1 color errors $\leq$0.05.  With this large of a sample it is more
informative to analyze the color distribution by taking a MS fiducial and creating a histogram of the color residuals
in C-T1 versus T1.  The left panel of Figure 8 shows this color distribution with bin sizes of 0.02, and we 
see that there is a significant sample of broadly distributed stars in the red wing that are not seen in the blue wing.  
A KMM mixture modeling test (Ashman et al. 1994) indicates that the probability this distribution is only a single 
Gaussian and not a bimodal distribution is essentially null.  Initial attempts to fit these two possible populations 
using two offset Gaussians of equal sigma were not satisfactory.  While the central peak is fit well by 
a Gaussian with a $\sigma$ of 0.033, attempting to fit the redder population with a similarly narrow $\sigma$ does a 
poor job of matching both the extended red wing and the blue wing.

Based on the analysis of synthetic CNO variations and abundances by Carretta et al. (2011b) and Ca11, they 
find that in colors involving Stromgren u the two RGB populations in NGC 1851 are not distinct sequences, 
similar to our C observations.  In these papers they argue that the redder population is significantly 
broader in color than the blue population and the two populations are heavily overlapped, where the full redder
population extends nearly as blue as the primary population.  This suggests that our observed redder population 
may also only be the reddest wing of a broadly distributed population.  Our synthetic analysis of the 
effects of CNO variations on the C magnitude, which will be discussed in detail in Cummings et al. (in prep.), 
shows the C filter also creates two heavily overlapping populations with a broader population extending well 
into the red.  The left panel of Figure 8 shows our fit of the color distribution when assuming
these population characteristics, with a large but narrow population with $\sigma$=0.031 and a smaller but 
significantly broader population with $\sigma$=0.075.  This second population extends as blue as the first but also 
extends significantly redder creating the second population in the red wing.  A clear advantage of this fit is 
that both the extremely blue and red stars are in agreement with the two Gaussians, which was not the case when 
using two slightly overlapping Gaussians of equal $\sigma$.  For clarification this broad second population
has partly been broadened further in this color distribution because its relative color difference between
the primary population increases in the fainter (cooler) stars, so at any given magnitude its breadth will be 
smaller than 0.075 but still more significant than the narrow population with $\sigma$ of only 0.031.
	
The peak heights and $\sigma$ values of the blue and red distribution Gaussians are 556 and 0.031 
and 99 and 0.075, respectively, giving that the broad and red MS is 30.1\% of the total population.  This is
comparable to the ratio of 27.9\% found for the percentage of blue HB to all HB stars, which are two far more
reliably separated populations.  This ratio is significantly larger than 
the population ratios found from Figure 3 and 5 of 13.6\% for the red RGB, 9.5\% for the red MS, and 
12.7\% for the faint SGB, but as suggested these smaller red populations analyzed are only the extreme red 
wing of the full broadly distributed second population.  For independent comparison, M08 found that their 
fainter SGB was $\sim$30\% of the total SGB population. Similarly, H09 found in their U-I observations
that the secondary red RGB and faint SGB branches were also $\sim$30\% of the total population. 

To further test these significantly overlapping Gaussian fits, we have reanalyzed in C-T1 versus T1 the RGB
from Figure 2 using an identical color distribution method.  The right panel of Figure 8 
shows the RGB color distribution for the 319 stars from 17.5$<$T1$<$14, and we can fit it quite well with two 
Gaussians very similar to those we used for the MS.  Again, the KMM test very strongly favors a bimodal
over a unimodal Gaussian distribution.  As discussed in Section 3, the color separation of the two RGB populations 
increases at brighter magnitudes, even more than that seen in the MS.  Therefore, even with significantly lower
color errors, the redder population requires an even broader Gaussian ($\sigma$ of 0.09 instead of 0.075). 
The two RGB populations are also more separated in color, with the Gaussian centers separated by 0.044 in the 
MS but by 0.064 in the RGB.  Consistent with this difference, we will discuss in detail in Cummings et al. (in 
prep.) that CNO variations will cause differences in C magnitude in both the MS and RGB, but the differences is 
greater in the typically cooler RGB stars.  Lastly, the peak heights and $\sigma$'s of the blue and red 
distributions are 67.5 and 0.036, and 12 and 0.09, respectively, giving that the red RGB is 30.8\% of the total 
RGB population.  This is again in strong agreement with that found for the two MS populations and the two HB 
populations.  This agreement adds further strength to characterizing the two populations with this method.

\subsection{Comparisons to Previous Analysis \& Multiple Main Sequences:}
	
In M08, where the split SGB was first discovered in NGC 1851, they also looked for evidence 
of MPs in the MS.  In contrast to what we have found, M08 did not find a second or even broadened MS, but they 
primarily searched for a clear split in the MS like that observed in NGC 2808 (Piotto et al. 2007) rather than 
two heavily overlapping populations such as we have found evidence for in our observations.  Furthermore, 
there are several key differences between their analysis and ours that may explain why they would not have 
found evidence for a second MS population.  First, they analyzed the MS with F336W and F814W (Johnson U and 
I) filters, but they 
did their analysis using only U-I versus U.  Plotting versus a UV magnitude, as we discussed in 
Section 4.1, diminishes the color separation of the two populations because the second population will 
be shifted both redward and fainter, partly parallel with the MS.  While there remains
signatures of a moderate redder population in their CMD (see Figure 4 of M08), they 
do not appear as significant as those we detect.  M08 do not mention these redder stars and they possibly 
assumed that they were a standard binary sequence.  Second, in their color distributions they 
removed all stars that were more than 4$\sigma$ from the center of the distribution, likely removing many of these extreme 
red stars.  Third, their U-I color distribution is also quite broad, giving a large $\sigma$ of 0.052 in their 
upper-MS stars and even greater $\sigma$ in their fainter MS.  In contrast, our primary
MS exhibits a narrower $\sigma$ of 0.031, suggesting that their color errors are more significant.  
M08 also acknowledge that their U magnitude errors are large, but they do not test in detail if 
their errors can fully explain the large $\sigma$.  All of these factors will have limited the 
significance of this second population.

Intrinsic differences between the U and C filters may also play an important role in causing this second MS to be
undetected.  In Cummings et al. (in prep.) we show that, assuming a constant C+N+O as found in V10,
variations in individual CNO abundances that are consistent with the spectroscopic analyses of NGC 1851 (Gratton et 
al. 2012a and Ca11) predict magnitude differences in C for upper-MS stars comparable to what we are observing 
here, and with no meaningful magnitude differences in R (T1) or I (T2).  When applying identical CNO variations to F336W
(U) we find that the predicted
U magnitude variations are comparable to the variations in C magnitude.  However, if we apply a moderate C and O 
anticorrelation, which due to the characteristics of U is important to consider, we find that 
the U magnitude variations will be relatively weaker ($\sim$80\%) than that predicted in the C filter.  It should 
also be noted that if the difference in C abundances across the two populations is increased, a possibility due to 
the limited number of red MS stars observed in Gratton et al. 2012a, the ability of the C filter to detect the 
two populations in the MS is greatly increased over the U filter.  Lastly, possible He and metallicity differences 
(Joo et al. 2013 and Ca11) may also play an important role by how the two populations differently affect U versus C.

MPs in globular cluster MSs have only been photometrically observed in a small number of clusters 
before, first with Omega Cen (Bedin et al. 2004) and NGC 2808 (Piotto et al. 2007), and more recently 
with 47 Tuc, NGC 6752, NGC 6397, and NGC 6441 (Anderson et al. 2009, Milone et al. 2010, 2012b,c, 
Bellini et al. 2013).  However, all of these multiple MSs have only been detected with HST photometry.  
Therefore, it is quite remarkable to detect a possible second MS using relatively little 
telescope time on a ground-based 1-meter telescope.  Another important difference with our second MS is that, 
other than in the very complex case of Omega Cen, these previously observed secondary sequences are bluer 
than the primary MS.  But it should be noted that the secondary RGB populations also 
are bluer in NGC 6752, NGC 6441, and 47 Tuc, the opposite of what we observe in the RGB of NGC 1851.  
Carretta et al. (2010) also find from spectroscopic analysis of NGC 6397 that its O-poor/Na-rich (photometrically 
redder) population also dominates.  Therefore, this suggests that in comparison to these other clusters NGC 1851 may 
have unique differences between its MPs and that the mechanism of their formation may also be different, 

\section{Multiple Red Horizontal Branches:}

Previous published photometry of the RHB for NGC 1851 has already shown it to be particularly 
broad in magnitude (e.g. Grundahl et al. 1999, M08, and H09).  In our observations shown in Figure 2 in C-T1 versus C 
and versus T1 we find evidence for a split in the RHB.  Further analysis of this split structure can 
be performed by analyzing where the stars of the sequences fall in 
a variety of CMDs.  Figure 9 shows all CMD combinations using our 3 filters with additional CMDs based on V.  
The CMD of C-T1 versus T1 has the most significant split.  Based on the C-T1 versus T1 diagram, we have 
separated the RHB into 2 groups: a brighter and redder group and a fainter and bluer group.  These identical color 
markers are applied to these same stars in all other CMDs shown, and we have included representative error 
bars determined by the median magnitude and color errors of only the stars we have colored.  

Quite strikingly, while the plots involving C-T1 show the clearest separated sequences, nearly all 
of the RHBs still show the 2 color groups with little overlap.  These results indicate that 
while we do not observe distinct and separated groups in T2 magnitude, the two RHB groups have a 
meaningful difference in this magnitude similar to their difference in T1.  The lack of more distinct 
T2 magnitude differences may solely be the result of the T2 errors being nearly double those of T1 for 
these stars.  The only combinations that do show significant overlap 
of the two groups are T1-T2 versus C and to a lesser extent versus V.  The overlap in color is
expected due to the consistency of the T1 and T2 magnitudes in these RHB stars, hence in the T1-T2 
color the two groups have no meaningful difference.  The overlap in magnitude suggests that for both C and 
V there is not as significant of a magnitude difference between the two groups.  It is unclear what 
could cause the RHB stars to have differing sets of T1 magnitudes and a 
significant spread in T2 magnitudes (if not distinct sets) at consistent C. However, these figures further suggest 
that at least the significant spread, if not separate sequences, in the RHB are a real feature in T1 and T2.  
While the BHB and the RHB already represent the two established populations in NGC 1851, does the structure in the 
RHB suggest further possible differences within its single population?

A split RHB has been seen before in Terzan 5, 47 Tuc, NGC 6440, NGC 6569, and NGC 6388 (Ferraro et al. 2009,
Milone et al. 2012b, Mauro et al. 2012, and Bellini et al. 2013).  The split in Terzan 5, NGC 6569, and NGC 
6388 were all observed using J and K filters.  Our T1-T2 versus T1 and T2 CMDs have similar characteristics to 
those observed in J and K: two clumps that have a moderate spread in color and are offset in magnitude.  
Similar to NGC 1851 the brighter clumps in both NGC 6440 and Terzan 5 are slightly redder than the faint 
clump, but in NGC 6569 there is no significant color difference between the two groups.  In NGC 1851
the RHB clumps have a T1-magnitude difference of $\sim$0.1, which is similar to the K-magnitude difference
observed in all other clusters besides Terzan 5, which has a striking $\sim$0.5 K-magnitude difference.
The split RHBs in 47 Tuc and NGC 6388 were observed differently by using combinations of various UV filters
and F435W, F606W, and F814W (the last 3 of which are comparable to B, V, and I).  The RHB structure in these 
filters also show similar characteristics to ours, and again this suggests that these split sequences of stars 
are not meaningfully different in U but are in I.  However, we should note that in 47 Tuc, Milone et al. (2012b) 
show that its double RHB is also visible but not clear in F275W-F336W.

Two groups have recently modeled the HB in NGC 1851, which provides us important comparisons.  First, 
Kunder et al. (2013) performed HB synthesis based on the abundances described in 
Gratton et al. (2012b), and they were able to create a very broad RHB in V-I versus V but with no split.  In 
our similar V-T1 versus V diagram we also observe a broad RHB with no clear split, but Figure 9 demonstrates 
that in this color the brighter extension of the RHB is composed of the brighter of the two split sequences 
we observe in C-T1.  Kunder et al. (2013) considered in their synthesis the Ba-rich, Na-rich, and O-poor RHB 
subpopulation ($\sim$10\%) from Gratton et al. (2012b), but they find that it is not related to the brighter RHB 
extension, which suggests abundance differences are not the key here.  Second, Joo \& Lee (2013) did full 
cluster models for two populations in NGC 1851, and they were able to reproduce the general photometric 
characteristics of NGC 1851.  As already well established, their two HB populations recreate the distinct RHB 
(primary population) and BHB (second population) with no overlap, but within the single RHB population they 
created in V-I versus V the heavily broadened RHB with even a weak possible split.  Their RHB structure was 
created through variations in stellar masses and mass-loss rates.

Consistent with this RHB split occurring in NGC 1851 without meaningful abundance variations, the HB abundances 
from Gratton et al. (2012b) show 
that the RHB stars have no meaningful spread in Fe or Ca.  In contrast to this, the two RHBs in Terzan 
5 show a significant $\sim$0.5 dex difference in [Fe/H] (Ferraro et al. 2009; Origlia et al. 2011) and a $\sim$0.3 
dex difference in [$\alpha$/Fe].  These abundance differences, combined with a possible large age 
difference, are likely the reason for the far more significant magnitude difference observed between its two RHBs. 
At a less significant magnitude, the metallicity analysis of NGC 6569 by Valenti et al. (2011) suggests that 
overall it has a bimodal metallicity with a difference of $\sim$0.08 dex, but this is based only on a very limited 
sample of 6 RGB stars and is not a direct comparison of stars in the two RHB groups.  

All of these split RHBs photometrically show many similar characteristics, in particular
by being most prominent in red and IR photometry.  Additionally, their recent discovery in
a number of GCs, including in already heavily studied clusters, suggests that it may be a more common
feature than previously thought.  Searches for more of these split RHBs and detailed analyses for any abundance
differences between the two clumps, as well as further understanding of the effects of mass loss, will be 
necessary to provide a complete picture for the cause or possibly multiple causes of this phenomenon.

\section{Radial Distributions:}

A useful method to analyze the potential formation mechanisms for the MPs in NGC 1851 is to see if they have differing 
radial distributions.  Zoccali et al. (2009) have argued that the two different RGB branches have a differing radial 
distribution, with the redder RGB population being more centrally concentrated, but both Milone et al. (2009) and
Olszewski et al. (2009) have found no evidence for this.  The spectroscopic analysis of Carretta et al. (2010), 
in contrast, presents evidence for variations in radial distribution based on their metal-poor and metal-rich 
populations.  In Ca11 they discuss their disagreement with the conclusions of Milone et al. (2009) and how the 
Milone et al. analysis methods may have prevented them from observing a statistically meaningful difference.  If the 
two populations have a differing radial distribution, this is consistent with the cluster having two star formation 
epochs, with an initial more extended population and a subsequent polluted population forming more centrally concentrated 
(D'Ercole et al. 2008).  However, this only represents the initial radial distributions and the 
difference would be slowly washed out as the cluster dynamically relaxes.  

Before analyzing our data for potential differences in radial distribution, we must consider possible 
effects of our zero-point photometric corrections (see Section 2).  In colors involving C a more 
centrally concentrated red population would create a redder average color near the core in comparison 
to stars at larger radii in both the MS and RGB.  This could possibly affect our spatial corrections in C.  
In contrast, this would have no effect on the comparisons of the clearly separated red and blue HB, 
and this would not affect our corrections of either T1 or T2.  Our analysis of the spatial correction in 
C with respect to V shows that while in total it is not radially dependent, it does have a 
radial component centered near the core of the cluster.  However, this component shows the core was
too blue rather than too red in C-V, and consistent radial components are also found in both T1 and 
T2.  This suggests that it is not an effect introduced by the two populations having differing 
radial distributions, or it would not similarly be seen in T1 and T2.  
These radial components observed are more suggestive of the effects of increasing 
crowding at decreasing radii, which as seen will affect all 3 filters and on average artificially 
brighten the crowded stars.  Therefore, these photometric corrections will not affect our analysis 
of the radial distributions for these populations.

For the radial distributions based on the two RGB branches, we have selected the blue and the red RGBs from 
14$<$T1$<$17.5 (see right-panel of Figure 7).  We have defined all stars with $\delta (C-T_1)$
from 0.06 to 0.30 as the red RGB (53 stars) and all stars from -0.075 to 0.02 as the blue RGB
(206 stars). This constrained color range helps to avoid regions with heavy population overlap.  
A KS test between these two RGB samples shows that they do not have a meaningful difference in radial 
distribution (p-value=0.55).  A similar comparison of the radial distributions of the blue HB (36 stars) 
and the red HB (93 stars), which are more reliably separated populations, yields that there also is no 
meaningful difference between the radial distributions (p-value=0.394).  While the small number of 
available stars in the RGB and HB limits this analysis, it should be noted that combining their samples 
still gives no meaningful difference in their radial distributions.

Use of the larger MS sample (see left-panel of Figure 7)
will greatly increase the number of available stars.  To create a smoothly-varying radial profile we will remove 
the small number of uncut MS stars that are within 2 arcminutes of the center.  To reliably assign stars to 
the two populations we will first avoid the regions with significant photometric overlap and define all stars 
with $\delta (C-T_1)$ from -0.035 to 0.0 as the blue population (919 stars) and all stars from 0.085 to 0.30 
as the red population (257 stars).  The upper panel of Figure 10 compares these samples
and we find at a significant level (p-value=0.0) that their radial distributions are distinct, with the red 
population being more centrally concentrated. This radial comparison for the two MS populations also 
solidifies our supposition that the red MS is composed of cluster stars and is not a contaminating sample of 
field stars, since the field-star spatial distribution would not significantly vary.  As a further test, we 
compare the radial distribution of this same blue MS population to a sample of redder stars from 0.02 to 
0.04 (424 stars).  Even though this second sample is redder, Figure 7 shows that it is still heavily 
dominated by the primary blue population.  In the lower panel of Figure 10 we show that these two blue MS 
samples show no significant difference in 
their radial distribution (p-value 0.371).  This indicates that our data is not affected by a systematic
color shift to the red at decreasing radii, which would similarly create our result in the upper panel of
Figure 10 but similarly create a higher central concentration for this redder wing of the blue MS.  This 
further strengthens the significance of our two population analysis in the MS and that the red MS has a
higher central concentration.

We can also analyze the radial distributions of the two groups we have observed in the RHB, and we find that 
the brighter and redder RHB (43 stars) does not show a meaningful difference (p-value=0.346) in comparison 
to the fainter and bluer RHB group (51 stars).  Similarly, the two RHB groups
in both NGC 6440 and NGC 6569 were not found to have differing radial distributions (Mauro et al. 2012).  
In contrast to this, the brighter group of RHB stars in both Terzan 5 and Tuc 47 were found to be more 
centrally concentrated (Ferraro et al. 2009, Milone et al. 2012b).  The reasons for the 
radial differences between the two RHB groups being observed in only some clusters may be 
another indication of different mechanisms for the formation of the two RHB groups.  For example, 
the large abundance differences and radial distribution differences observed in Terzan 5 may further indicate that 
the two RHBs are two distinct populations being observed.  In NGC 1851 its two populations are already 
represented by the BHB and the RHB.  Its two RHB sequences likely do not have further abundance differences 
and are only composed of one population.

Considering all of these radial distributions, we can argue there is a higher central 
concentration for the redder population only for the MS stars.  For the other evolutionary stages, there is
no strong evidence.  This may be the result of the smaller numbers (1230 for the MS versus 259 for 
the RGB and only 129 for the HB), but it most likely is indicative of other issues.  For example, the 
binaries that we have shown to be a minor component of this red MS also clearly have a higher concentration
in the core (see Milone et al. 2012a).  Are these binaries biasing our radial distribution analysis in the MS?
Possibly, but because they are a relatively small number of binaries, this likely is not at a level 
significant enough to fully explain our observations.  The lack of detection in the RGB and HB may also be 
indicative of the effects of mass segregation/dynamical relaxation in this cluster, which will wash out 
initial differences in radial distributions for the two populations.  This is also illustrated in the higher 
concentration of binaries in the core.  Do the low-mass MS stars provide a more reliable tracer of the 
initial radial distributions of these two populations, while the more massive RGB and HB stars have all 
become heavily concentrated in the center resulting in a weakened, if not completely removed, initial 
difference in radial distribution?

\section{Conclusions and Future Work:}

We have shown that the Washington C filter that is very broad and specifically designed for CN/CH is a powerful 
tool for detecting MPs in GCs.  This provides a great advantage for photometrically studying MPs because the 
previous methods have either used extremely high precision HST photometry or inefficient ground-based UV 
photometry using the narrower and bluer Johnson U or Stromgren or Sloan u filter.  This has previously made 
studying MPs in GCs telescope time intensive and limited in depth.  MPs have been observed in the RGB and SGB 
of NGC 1851 before, which we have observed as well, but our data was obtained in a relatively short time on a 
1-meter telescope. In addition, our observations with the Washington C filter have provided strong evidence for MPs 
in the MS of NGC 1851.  This is the first case of MPs on the MS of a GC being uncovered via 
ground-based data, and the small aperture we used makes this doubly impressive.

With the high-precision T1 photometry we have evidence for a double sequence in T1 and possibly 
T2 in the RHB.  Previously observed split RHB's have shown many similar characteristics to those we 
have observed.  In some clusters it is believed the split is caused by the two RHB clumps having 
different metallicity and possibly age (e.g. Terzan 5), but in all other clusters
there remains no strong evidence for a metallicity difference within the RHB.  When the split is only moderate (i.e. a magnitude
difference of $\sim$0.1) and there is no meaningful metallicity or radial distribution difference they may solely 
be the result of variations in the mass-loss rates.

Detailed analysis of our two observed MSs in all available filter 
combinations strongly argues that the redder sequence is caused by a fainter C magnitude for this population.  This 
argues that these stars are representative of a second population 
and not a binary sequence. We further test the reality of MPs on the MS by comparing to a binary sequence created through
population synthesis, and find further distinction between our observations and a binary sequence.  However, in our 
analysis of color outliers in T1-T2 we found that a subsample 
($\sim$6-7\%) of the red population show characteristics consistent
with a binary population that have high-mass ratios.  In comparison to the total population this
is $\sim$0.65\%, which is very small but similar to the high-mass-ratio binary fraction of 0.8$\pm$0.3\% 
observed in the core of NGC 1851 (Milone et al. 2012a).  

Based on models for the photometric effects of varying CNO abundance (Carretta et al. 2011b; Cummings et al in 
prep.), we have been able to fit the color distribution of the two populations on the MS with a narrow blue 
population and a heavily
overlapping and broader population that is centered moderately redder.  The red wing of this broader population
extends well beyond the blue population creating the broadened MS and RGB and the clear signatures of a MP.  
Fitting the full width of the broad second population shows that it is $\sim$30\% of the full population in both 
the MS and RGB.  This is in remarkable agreement with the 27.7\% ratio found from comparison of the BHB to the RHB 
numbers, where the BHB is believed to be representative of the evolved redder population in the HB.  It is also consistent 
with the population ratios found by M08 and H09.

Lastly, whether there is a difference in radial distribution between these two populations remains a heavily
debated topic.  Independent analysis of our two photometric populations in all parts of the CMD show that in the
high-mass stars (but using relatively limited numbers) in both the RGB and HB there is no meaningful difference 
found in the radial distributions of the two populations.  However, in the low-mass MS stars there is a statistically 
significant difference in the radial distributions of the two populations, where the redder population is more 
centrally concentrated.  Is the differing result when comparing the high-mass and the low-mass stars a matter of 
small numbers, or is it indicative of the contaminating effects of binaries in the red MS that are known to have 
a higher central concentration, or does it indicate that the radial differences created during the formation 
of the cluster only remains in the low-mass stars and have already been dynamically washed out in the highest-mass stars? 
The binary explanation at least seems remote given the very low incidence of these stars.

These MPs in the MS, the SGB, and the RGB are all observed by their variations in C magnitude and consistent T1 or 
T2 magnitude.  The C magnitude variations are believed to be caused by variations in the strong CN, CH, and NH 
bands present in the C filter, suggesting a distinct difference between these populations
in C or N abundance, if not both.  This is consistent with previous spectroscopic analysis of NGC 1851 (e.g.
V10; Gratton et al. 2012a, Carretta et al. 2014).  We will discuss this in more detail in our following paper on 
NGC 1851 where we match available abundances to our photometry and analyze the possible correlations.
We will also synthetically analyze the effects of CNO, helium, and metallicity variations in the Washington and 
Johnson filter systems (Cummings et al. in prep).

Many questions remain about NGC 1851, and further models and abundance analyses of a wide variety of elements 
in all evolutionary stages of this cluster are needed to help create a complete 
and self-consistent picture for not only what the differences between these populations are but how these
populations formed.  Additionally, as with many questions in astronomy, analysis of MPs in as wide a variety of 
clusters as possible will be helpful in shedding light on these remaining questions.  With the considerable advantages that 
the Washington C filter provides for photometrically discovering and analyzing these MPs in massive star clusters, further 
Washington observations of a broad range of GCs will be very helpful in this endeavor. The 
efficiency of the system for investigating MPs means that studies of the host of Magellanic
Cloud clusters, with their range in age and metallicity parameter space not covered by Galactic GCs, are
within reach of 4 to 8-meter-class ground-based telescopes.  The inclusion of the F390W C filter in the
WFC3 on board HST opens up the exciting possibility of exploring MPs in M31 GCs.

The authors would like to acknowledge the sharing of data by Y. Momany. 
DG and SV gratefully acknowledge support from the Chilean
Centro de Astrofisica FONDAP BASAL PFB-06/2007 and the Chilean Centro
de Excelencia en Astrofisica y Tecnologias Afines (CATA). JC is thankful for
the financial support from Fondo GEMINI-CONICYT 32100008 and 
the support by the National Science Foundation (NSF) through grant AST-1211719.
SV gratefully acknowledges the support provided by FONDECYT N. 1130721

\begin{figure}[htp]
\begin{center}

\includegraphics[scale=0.63]{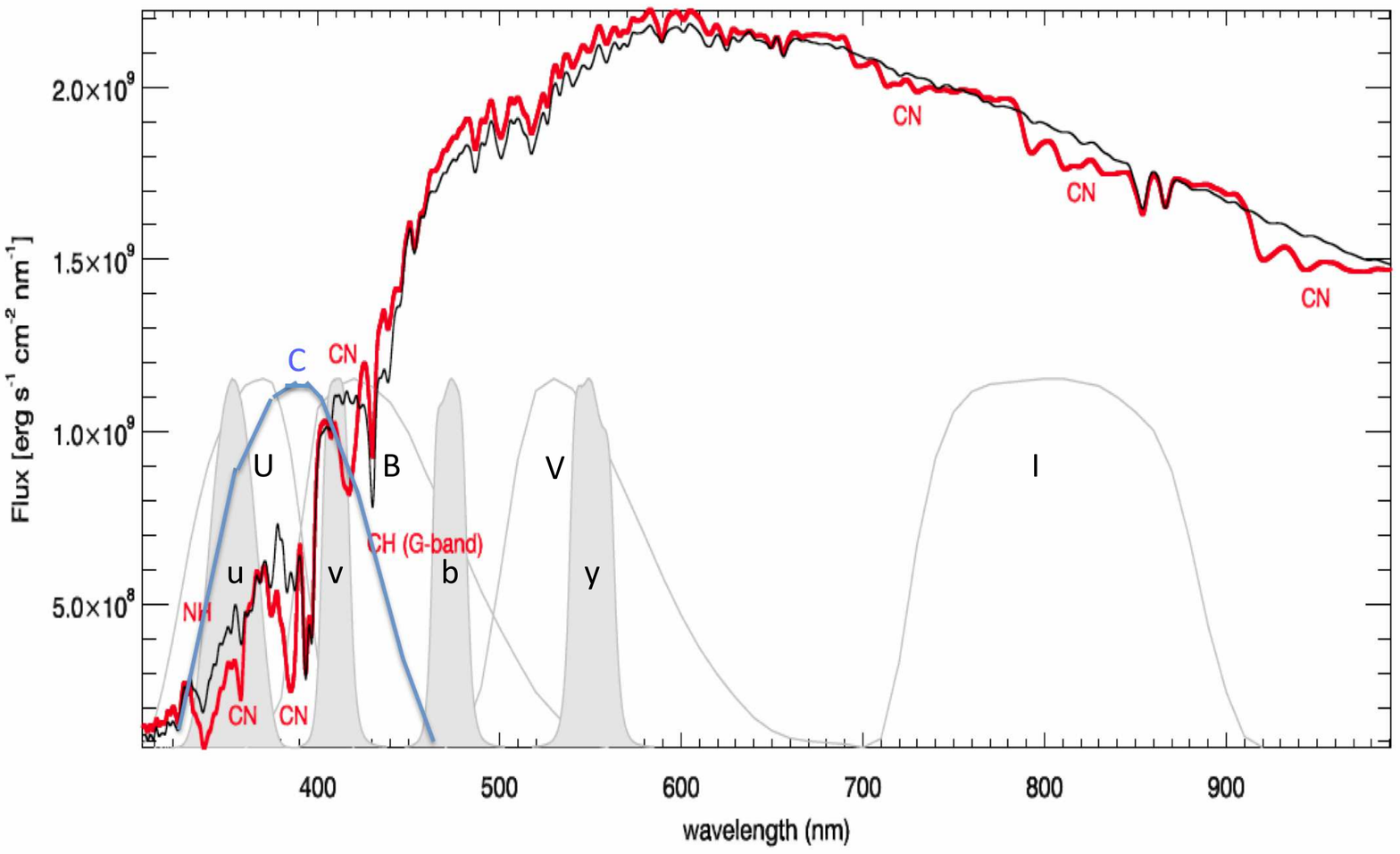}
\end{center}
\vspace{-2cm}
\caption{Synthetic spectra illustrating the large effects that variations in CNO can have on the molecular
band strengths in RGB stars and on selected filters.  Several of the important bands are labelled.  This figure has 
been created using Figure 4 of Sbordone et al. (2011), adding labels to all of the filter profiles and illustrating 
the Washington C-filter profile.  The filter response curves are only 
illustrative; e.g. the C filter peak response is significantly higher than the other UV filters. The black spectrum 
represents a typical first generation star with normal C, N, O, and Na while the red spectrum represents a typical second 
generation star with depleted C and O, and enhanced Na, and significantly enhanced N.  Overall, the total C+N+O is 
increased by 0.36 dex.}
\end{figure}
\clearpage

\begin{figure}
\centering   
\begin{overpic}[scale=0.87]{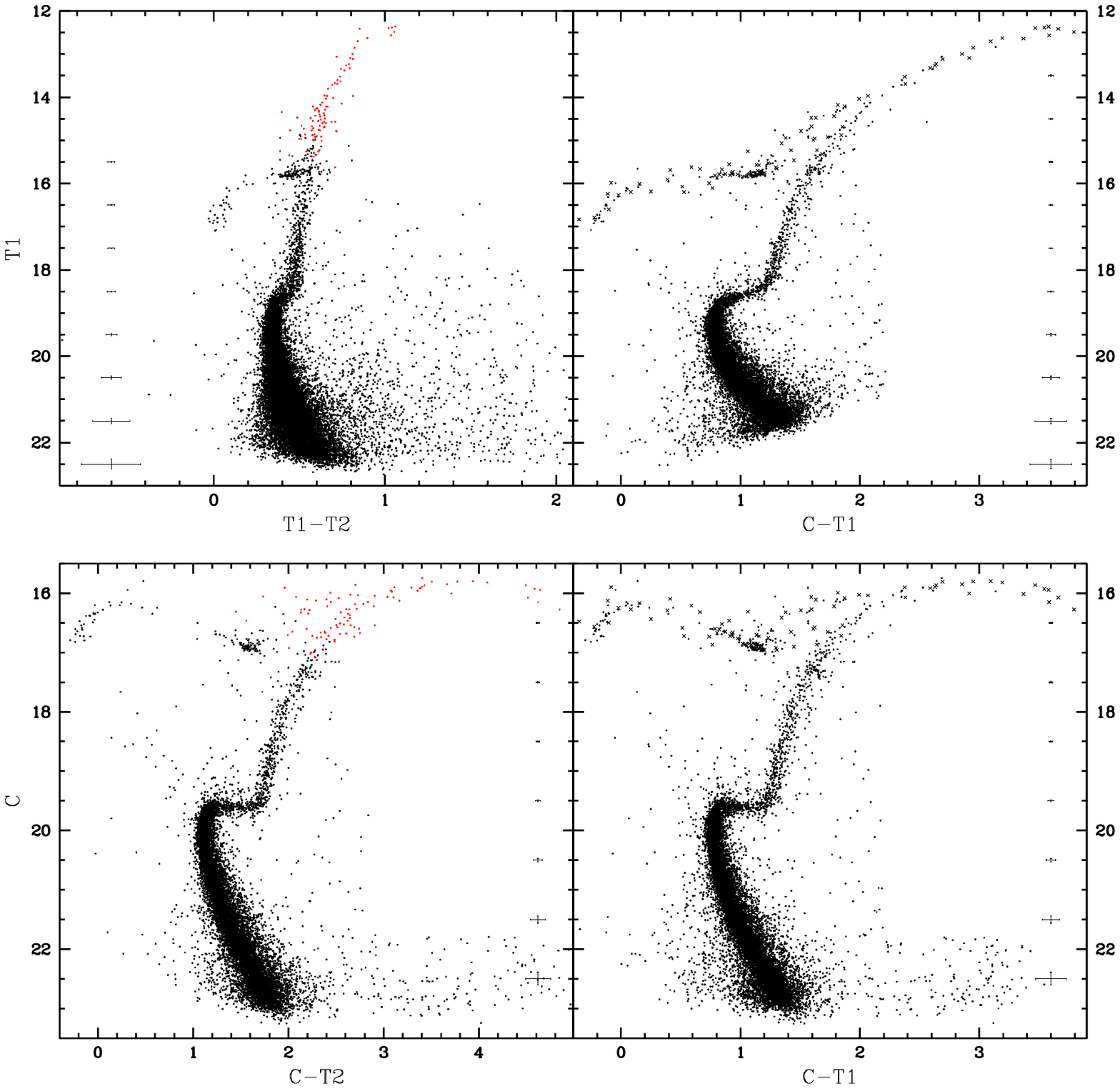}
\put(58.8,69.5){\includegraphics[scale=0.11]{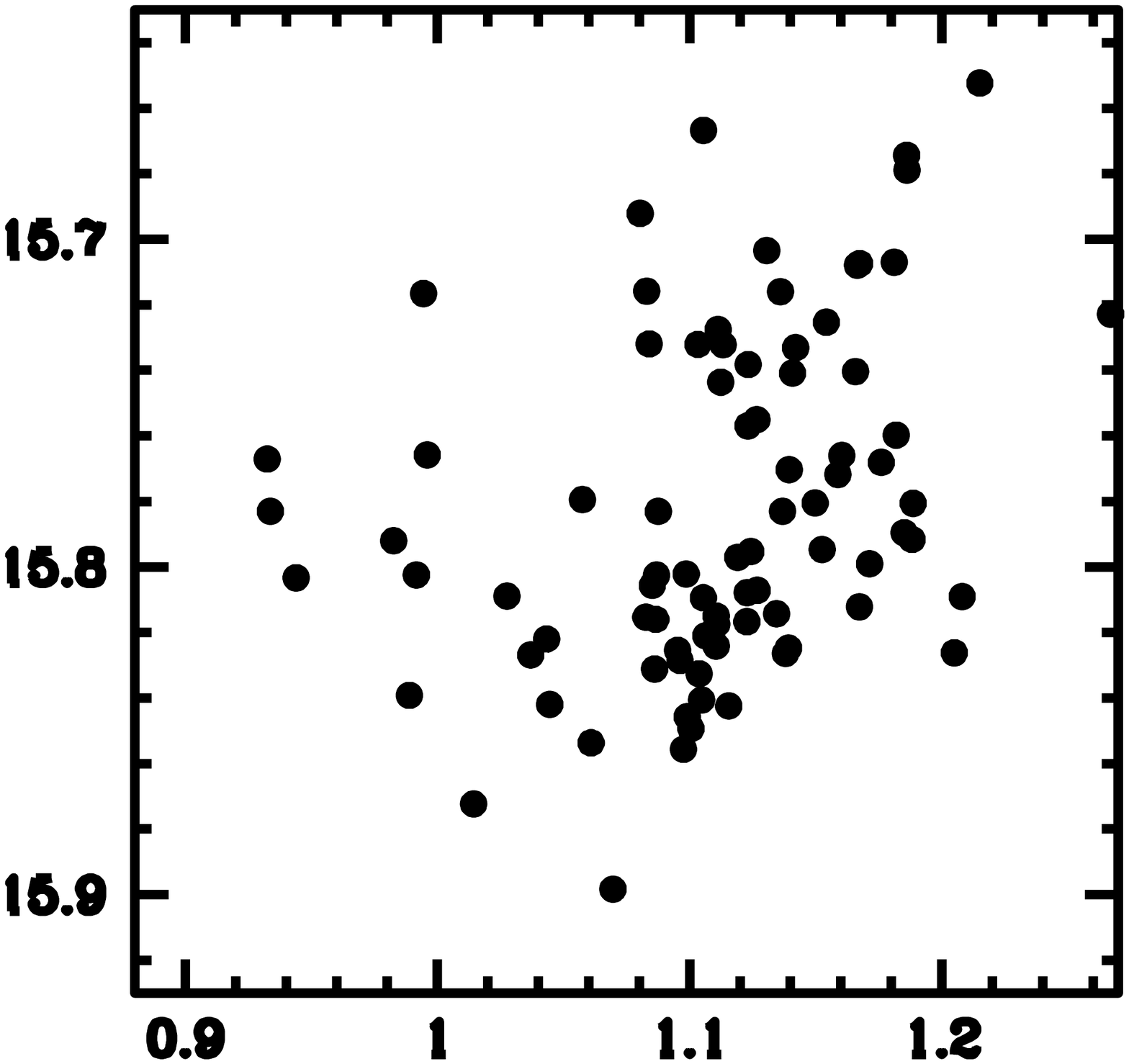}}  
\put(58.8,38.5){\includegraphics[scale=0.11]{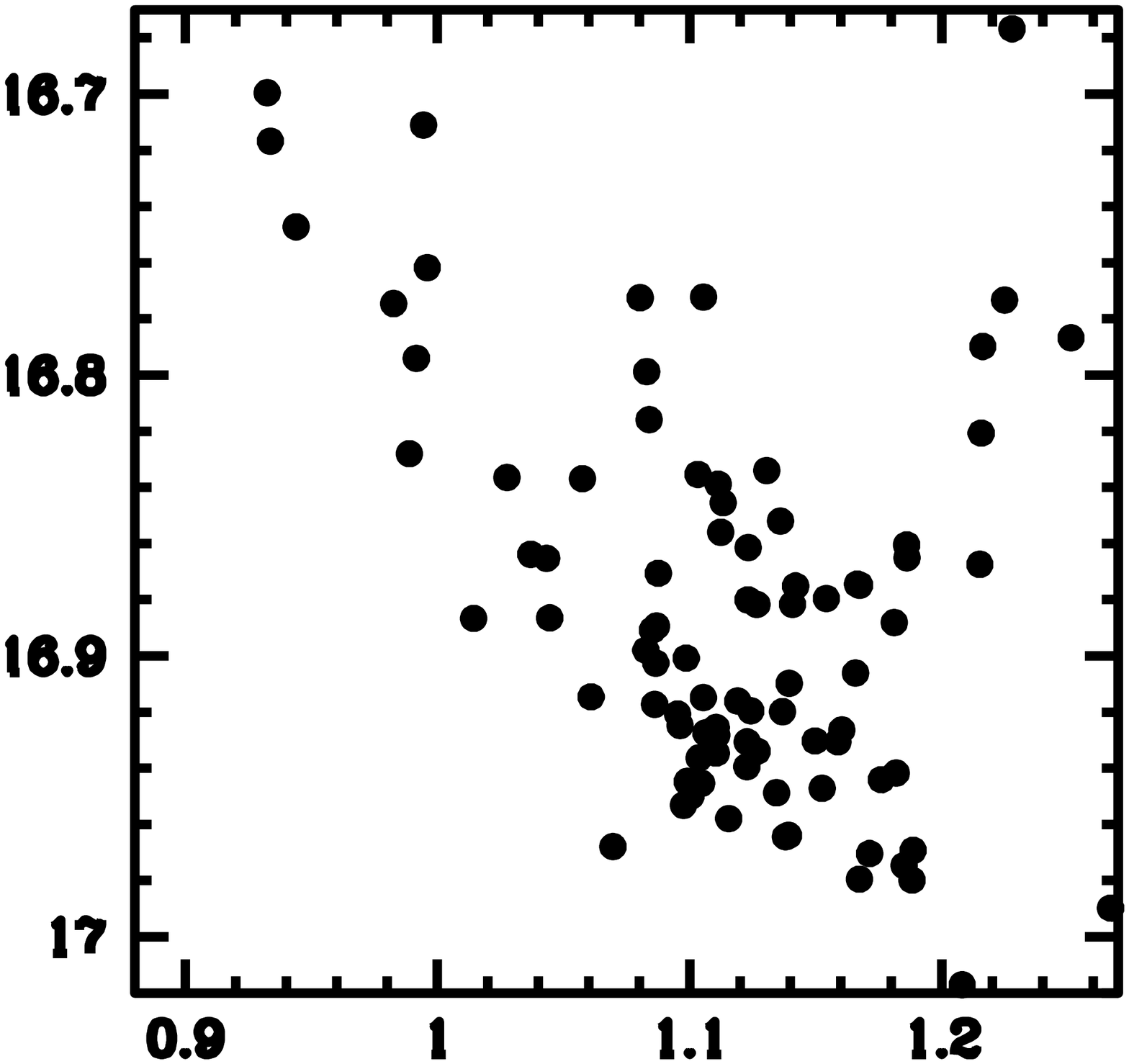}}
\put(58.8,60.5){\includegraphics[scale=0.11]{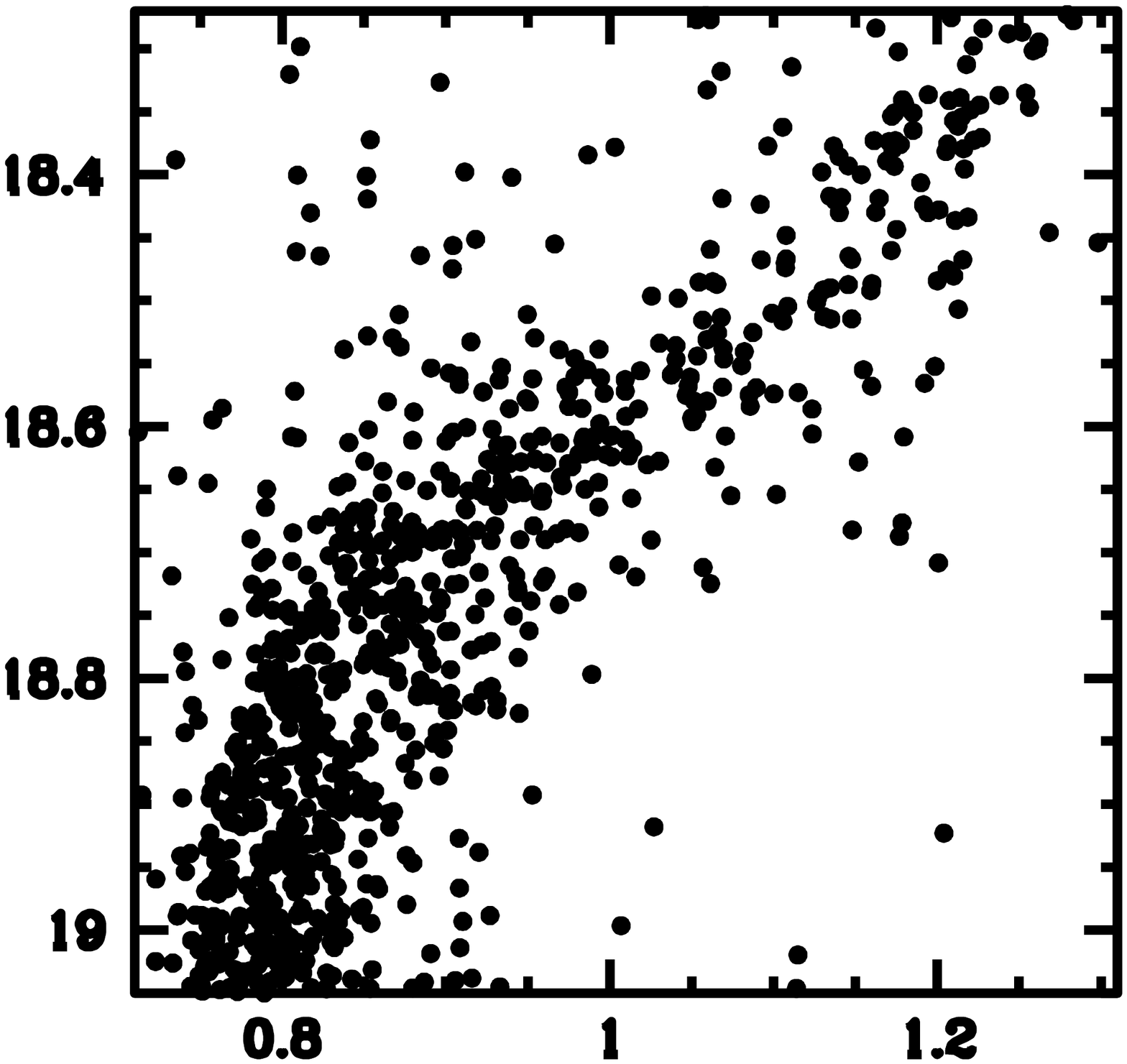}}  
\put(58.8,29.5){\includegraphics[scale=0.11]{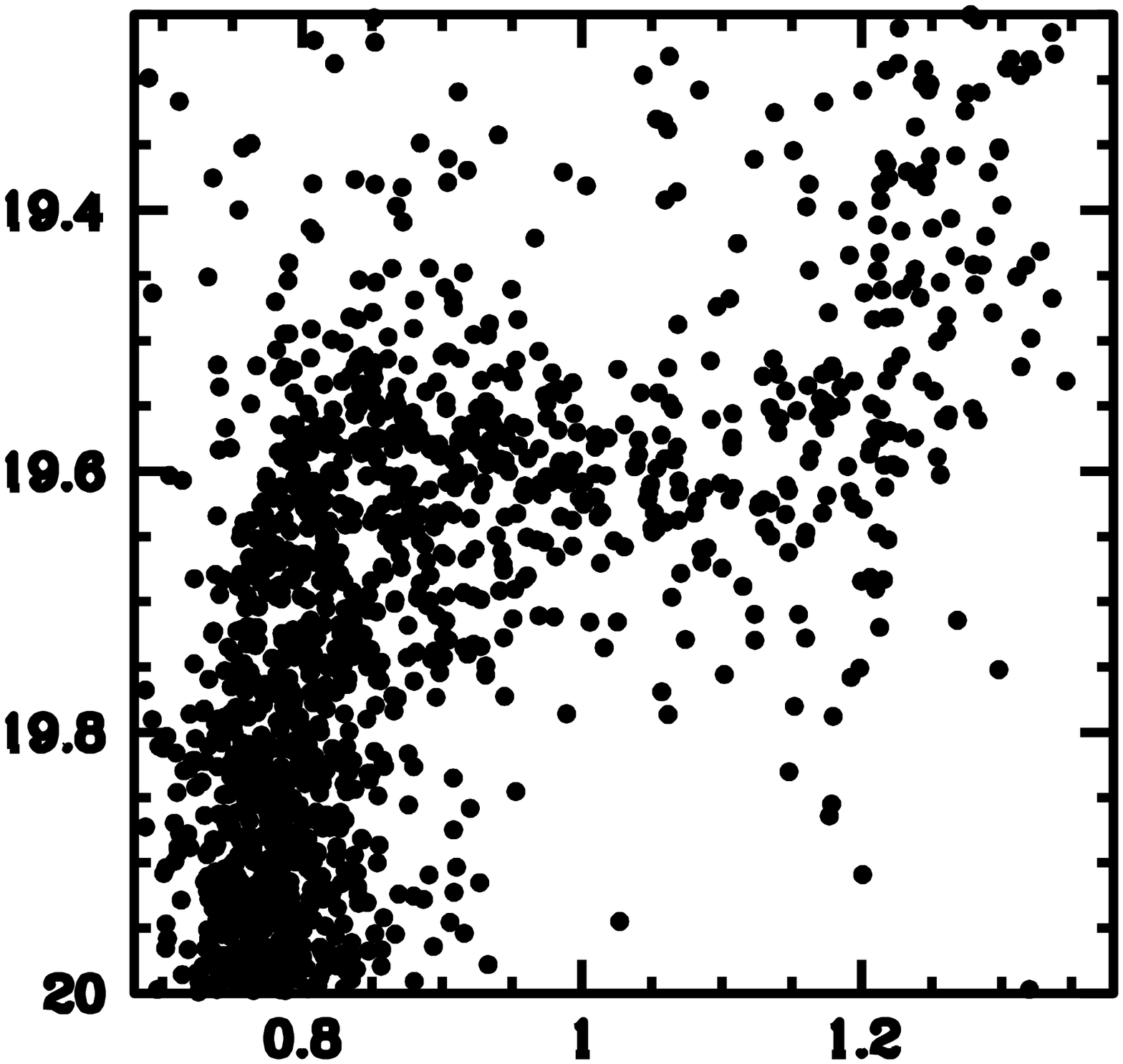}}
\put(28.5,38.5){\includegraphics[scale=0.11]{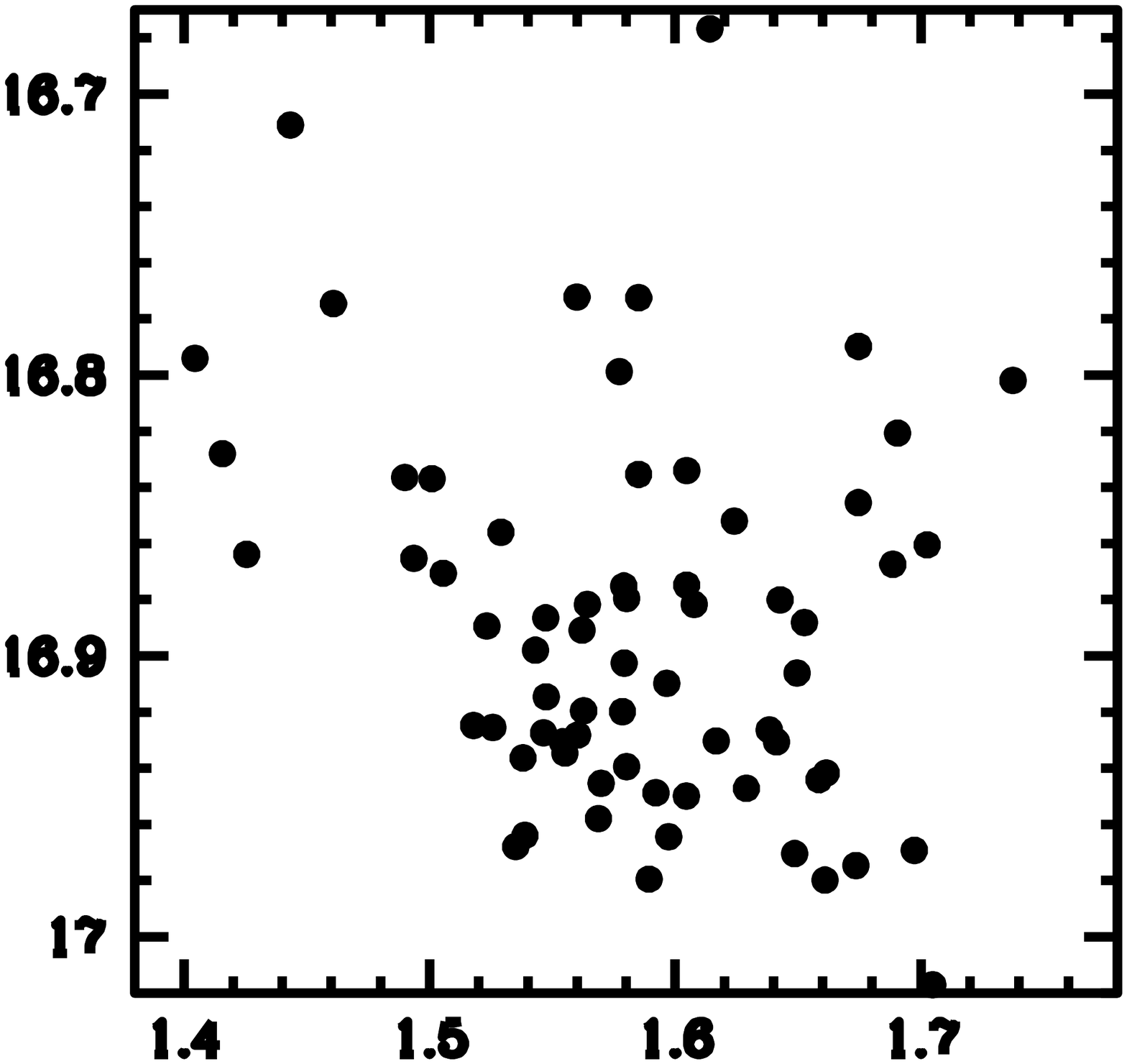}}
\put(28.5,29.5){\includegraphics[scale=0.11]{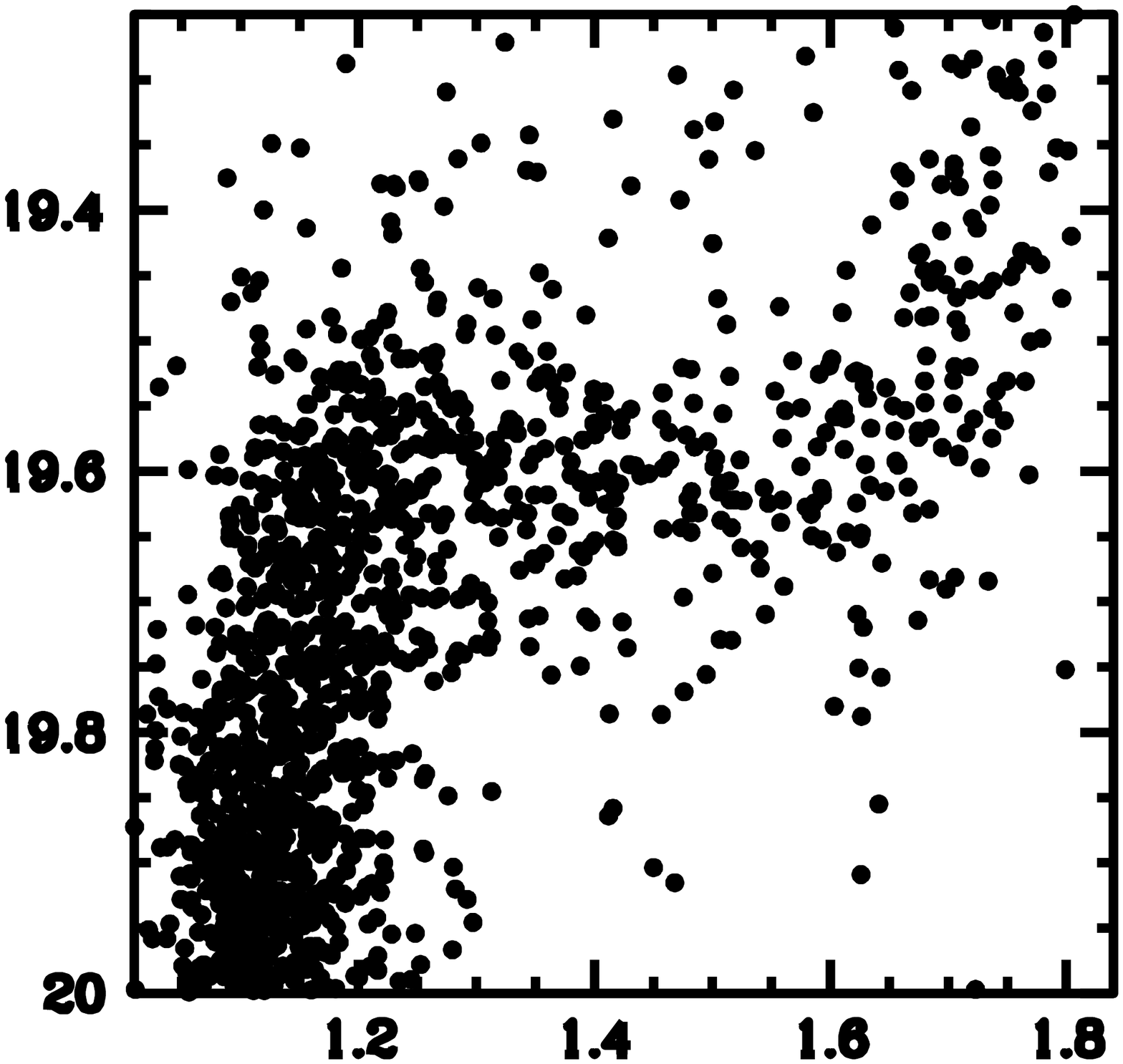}}
\end{overpic}
\vspace{-5.4cm}
\caption{Our Washington photometry of NGC 1851.  To account for the effects of crowding in the core of the cluster,
we require that for stars within 2 arcminutes of the center they have multiple observations in both filters and that
their resulting magnitude dispersions for their multiple measurements are less than 0.02 magnitudes in each filter.
Representative error bars are shown at each magnitude.  We expand our T2 magnitude range by converting the similar I
magnitudes to T2 and show these stars in red.  We expand both of our C and T1 magnitude range by matching
to the observations of Geisler \& Sarajedini (1999) and show these stars as x's.  The second redder population is 
observable in both the RGB and the SGB in all CMDs involving the C filter.  Insets are shown to more clearly
display the SGB populations, and shown in the RHB insets we can 
detect a split structure in both C-T1 versus C and T1 with a heavily broadened RHB in C-T2 versus C (See Section
5 for detailed discussion of the RHB).}
\end{figure}
\clearpage

\begin{figure}
\centering   
\begin{overpic}[scale=0.87]{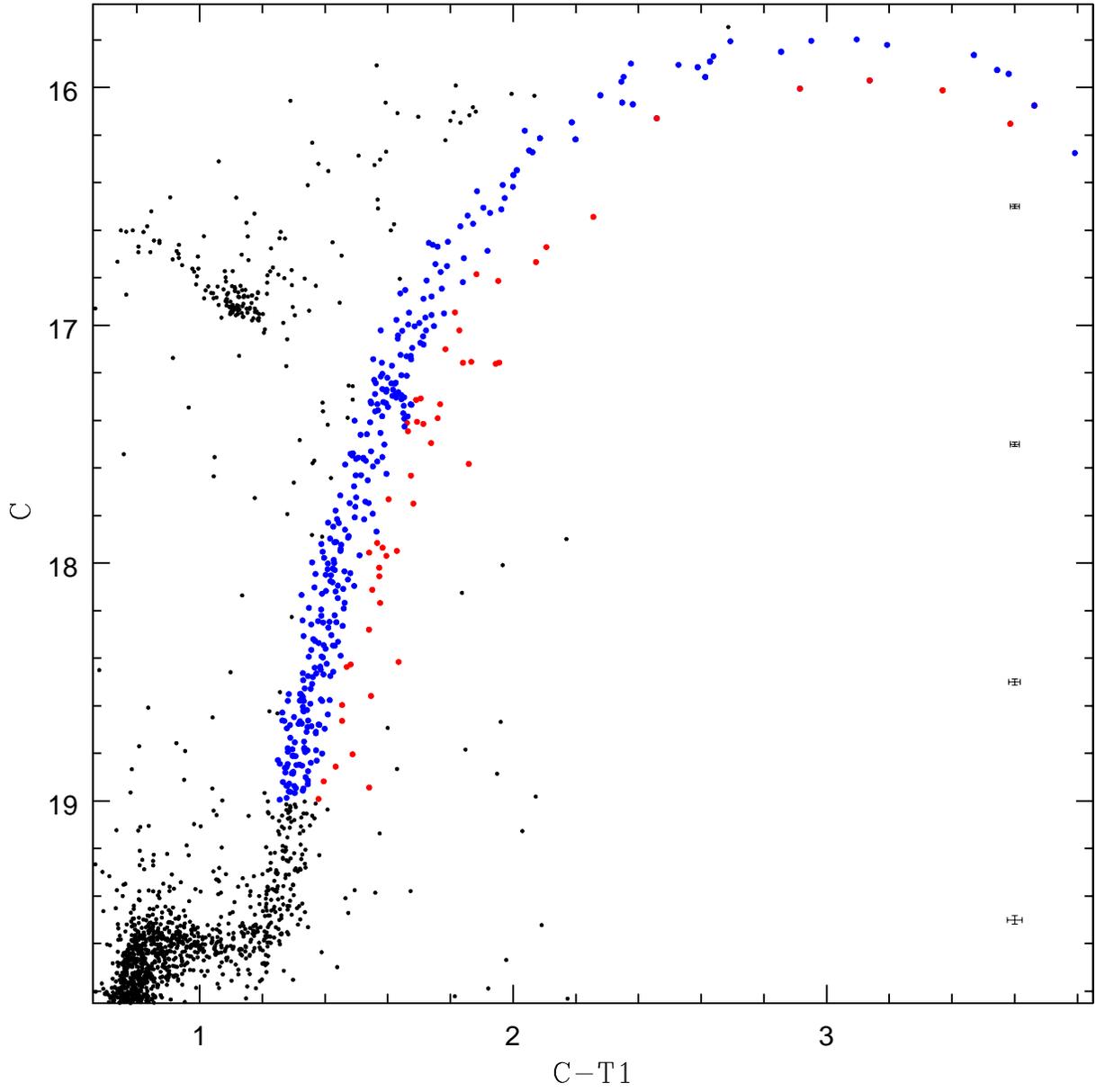}
\end{overpic}
\vspace{-5cm}
\caption{We focus on the SGB and RGB in C-T1 versus C, and to help clarify the branches we color the blue RGB
blue and the red RGB red.}
\end{figure}
\clearpage

\begin{figure}
\centering   
\begin{overpic}[scale=0.87]{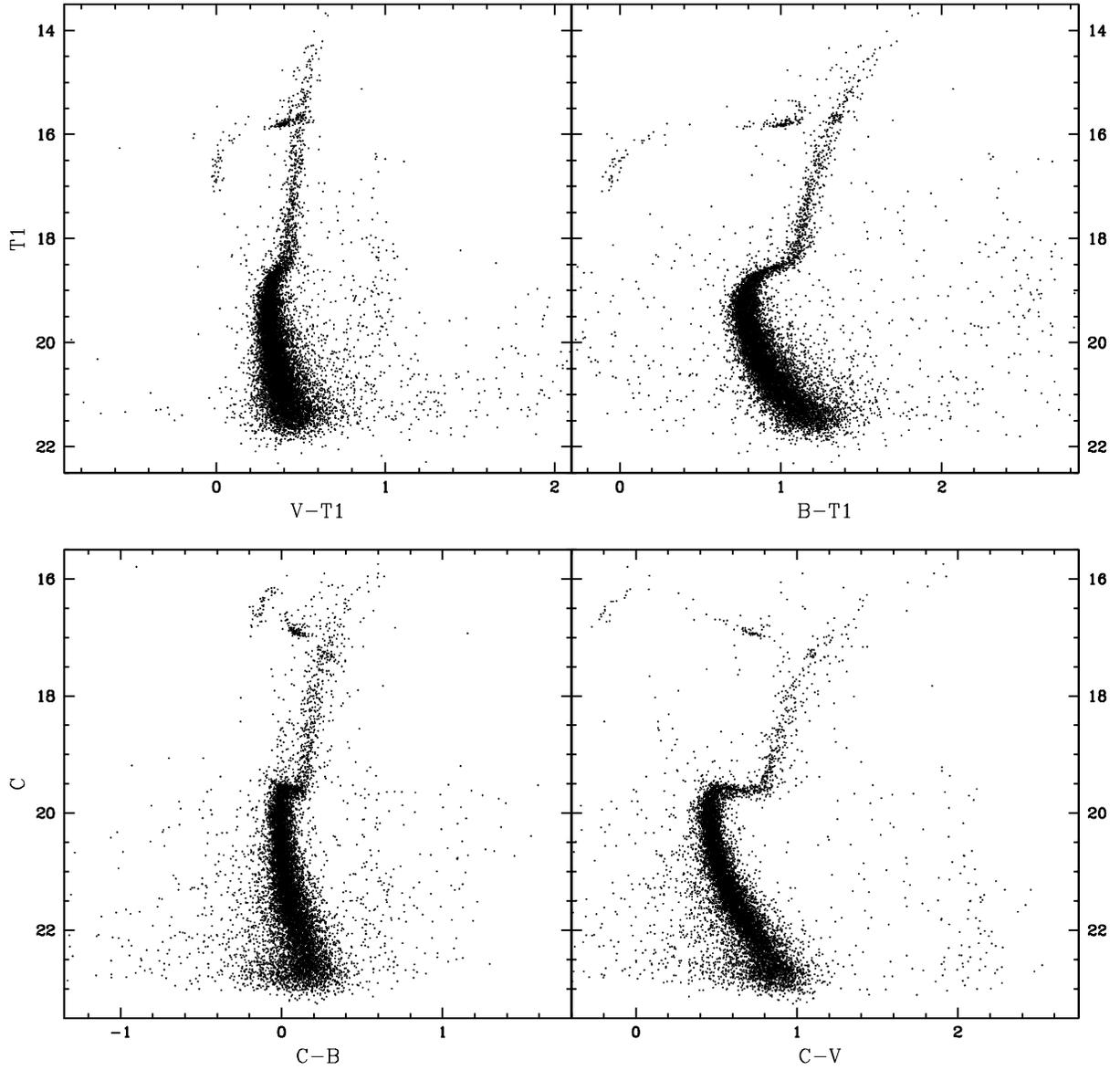}
\end{overpic}
\vspace{-5.4cm}
\caption{Additional CMDs that show the combinations of our Washington magnitudes and B and V.  These
show consistency with our CMDs in Figure 2, and also indicate that the two populations have moderately different
B magnitudes.  In Figure 1 we showed that the B filter also containing several CN and CH bands of importance
that would produce a change in B magnitude, but it has less bands than any of the UV filters producing a smaller 
change.}
\end{figure}
\clearpage

\begin{figure}[htp]
\begin{center}
\includegraphics[scale=0.87]{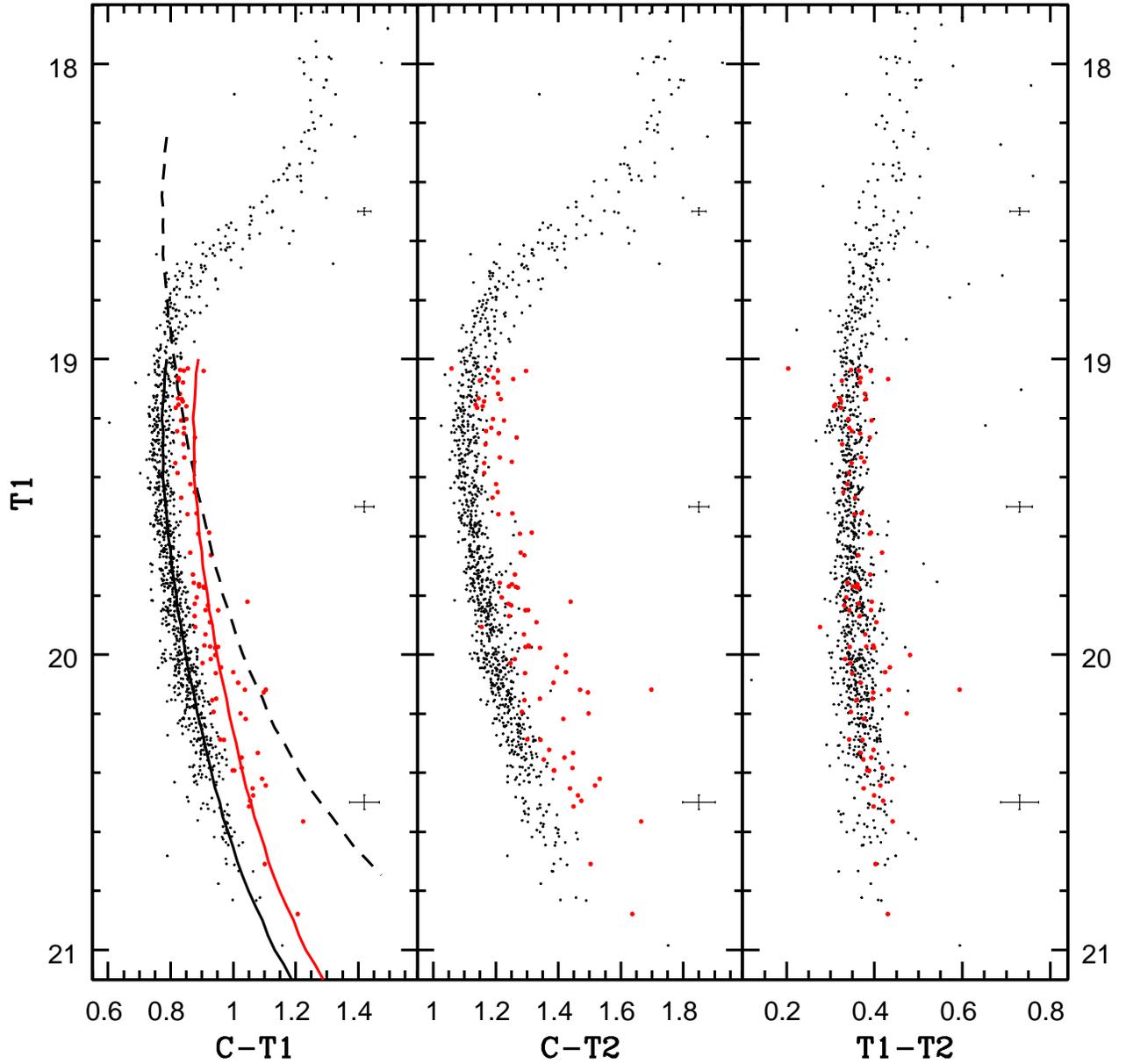}
\end{center}
\vspace{-5cm}
\caption{To focus more on the main sequence we use an inner and outer radius of 4 and 6.5 arcminutes and only 
stars with C-T1 color errors below 0.05.  Fainter than a T1 of 19 we have
selected stars belonging to the second redder sequence using C-T1 and they are marked as red.  To analyze how
these stars are distributed in the other colors we have similarly colored them in both C-T2 and T1-T2.
Additionally, in C-T1 we fit the MS fiducial shown in solid black, show the corresponding equal-mass
binary sequence in dashed black, and lastly show the MS fiducial shifted redward 0.1 in solid red.  The
second sequence is inconsistent with a binary sequence, and comparison to the solid-red line shows a
more comparable match but that the color difference also moderately increases in fainter (cooler) stars.}
\end{figure}
\clearpage

\begin{figure}[htp]
\begin{center}
\includegraphics[scale=0.87]{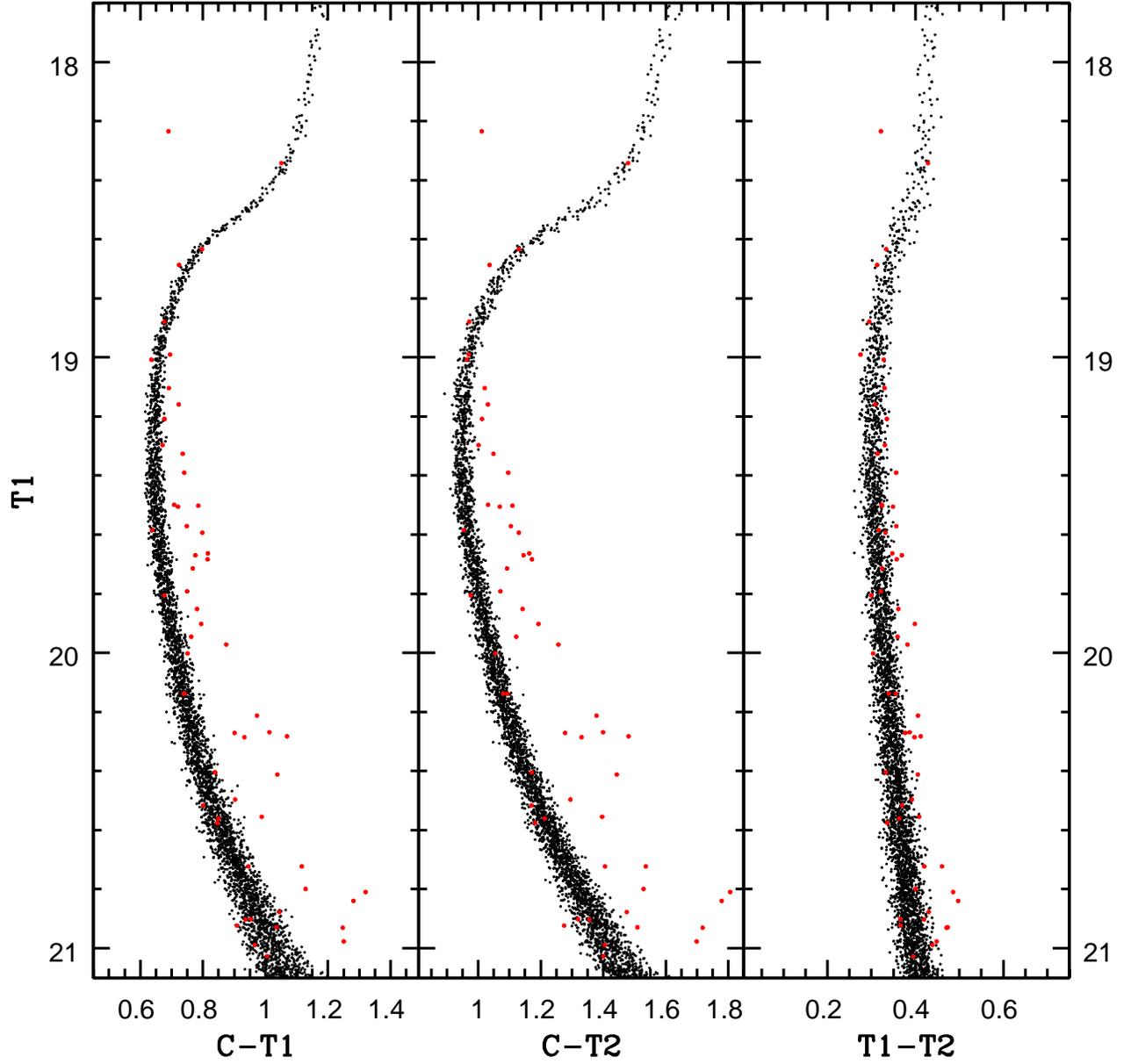}
\end{center}
\vspace{-5cm}
\caption{With population synthesis we have created a single population based on NGC 1851 with a moderate 
binary fraction of 3\%, which is larger than the binary fraction of 1.6\% determined by Milone et al.
(2012a).  We use the larger fraction for the purposes of making a clearer binary sequence.  The binaries with 
a mass fraction greater than 0.6 have been colored red in all of the CMDs.  This 
shows that the characteristics of a binary sequence and our second redder population are not consistent.}
\end{figure}
\clearpage

\begin{figure}[htp]
\begin{center}
\includegraphics[scale=0.87]{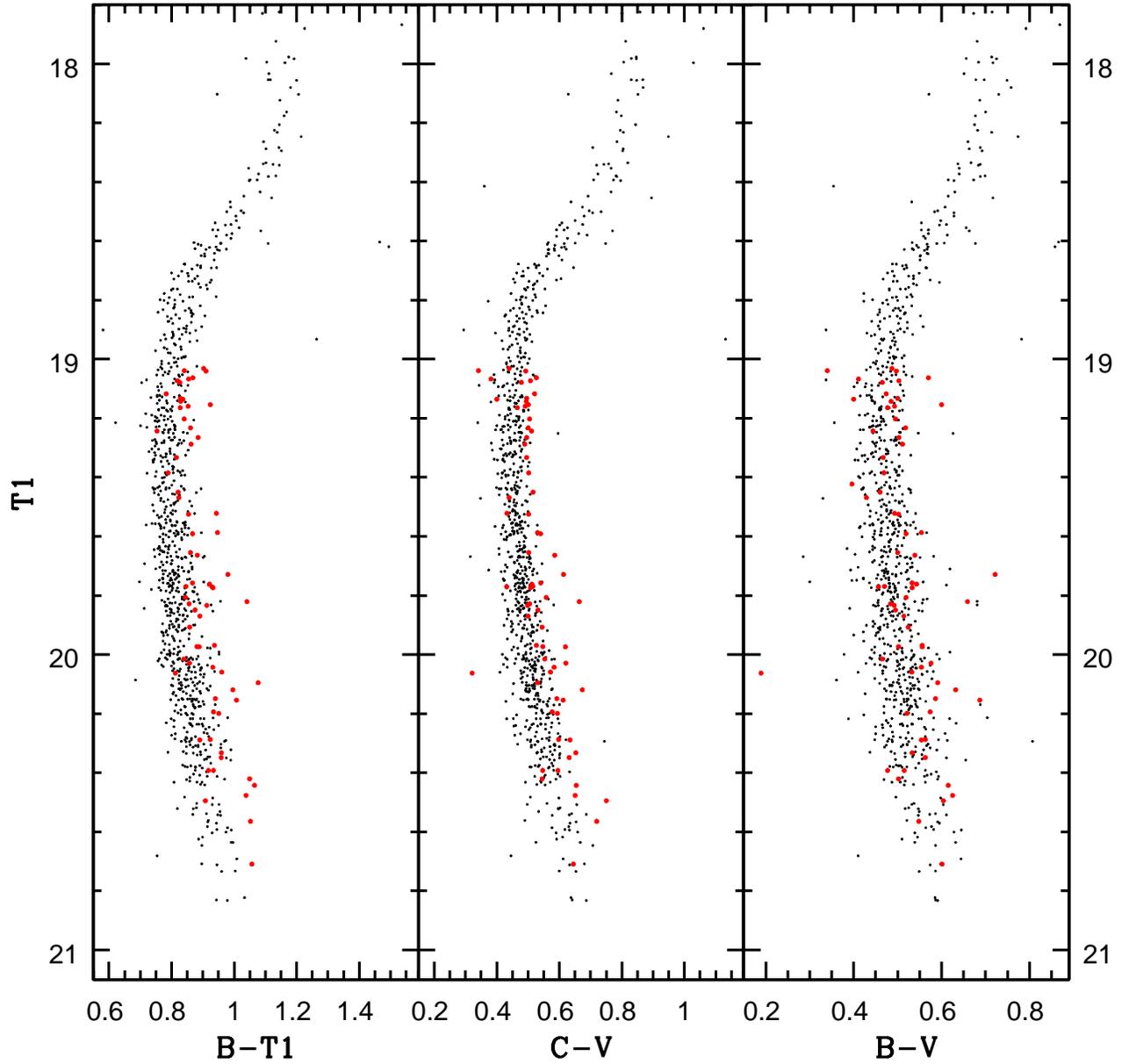}
\end{center}
\vspace{-5cm}
\caption{Using the same sample of stars in Figure 5, but comparing to additional colors based
on C, T1, B, and V magnitudes.  We again mark the red population stars from C-T1
with red data points.  We still find a second sequence of similar characteristics to that shown in Figure 5. 
Consistent with both Figure 1 and Figure 4 we find that there are important differences between the the populations
in B magnitude, but we also find evidence for minor differences in V magnitude for the two populations.}
\end{figure}
\clearpage

\begin{figure}[htp]
\begin{center}
\includegraphics[scale=0.9]{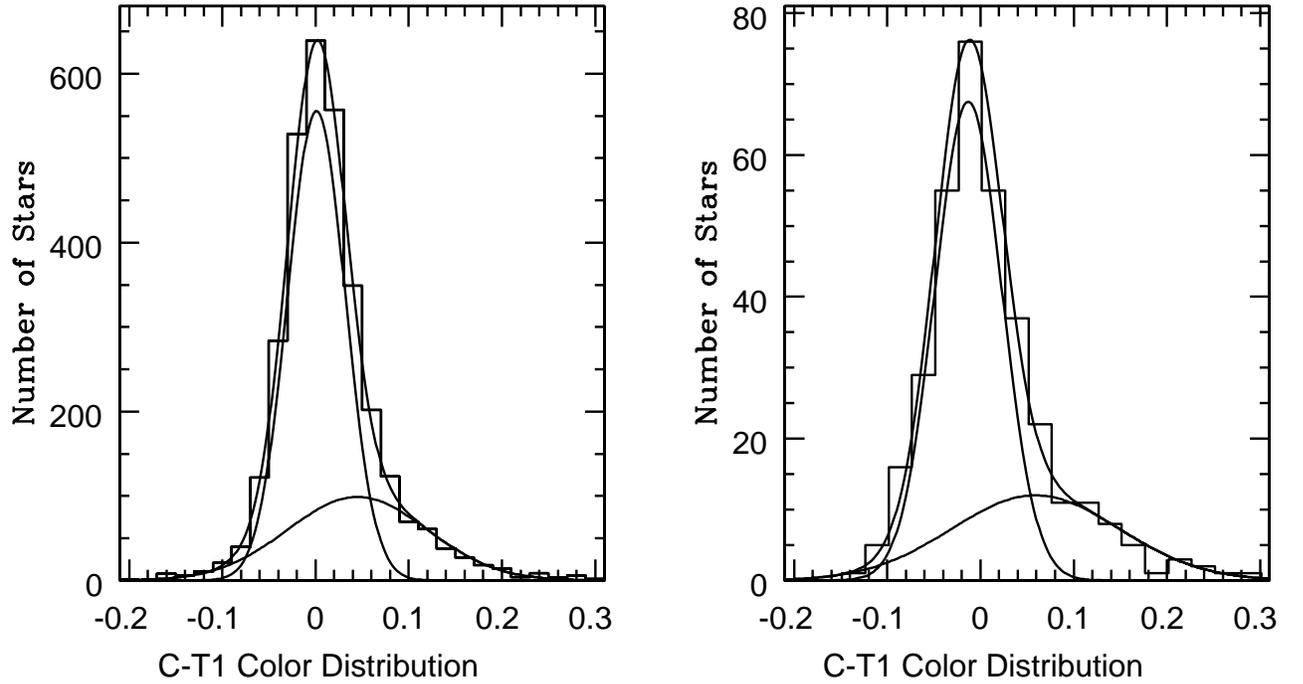}
\end{center}
\vspace{-5cm}
\caption{The C-T1 color distribution using all MS stars from 19$<$T1$<$21.5 with C-T1 color error $\leq$0.05 
and their difference from the MS fit shown in the left panel of Figure 5.  Two populations are fit by two 
Gaussian distributions, which have characteristics qualitatively based on the models of Carretta et al. 
(2011b).  The sum of the two Gaussians are also shown.  This reliably fits the full range of the color 
distribution, including the heavily extended wings.  The right panel shows the C-T1 color distribution 
of all of the RGB stars shown in Figure 2.  We have applied similar Gaussian fits as those shown in the MS.  
The Gaussian fits in both panels give two populations of $\sim$30 and 70\% ratios with a smaller population that 
is typically redder and covers a broader range in color.}
\end{figure}
\clearpage

\begin{figure}[htp]
\begin{center}
\includegraphics[scale=0.82]{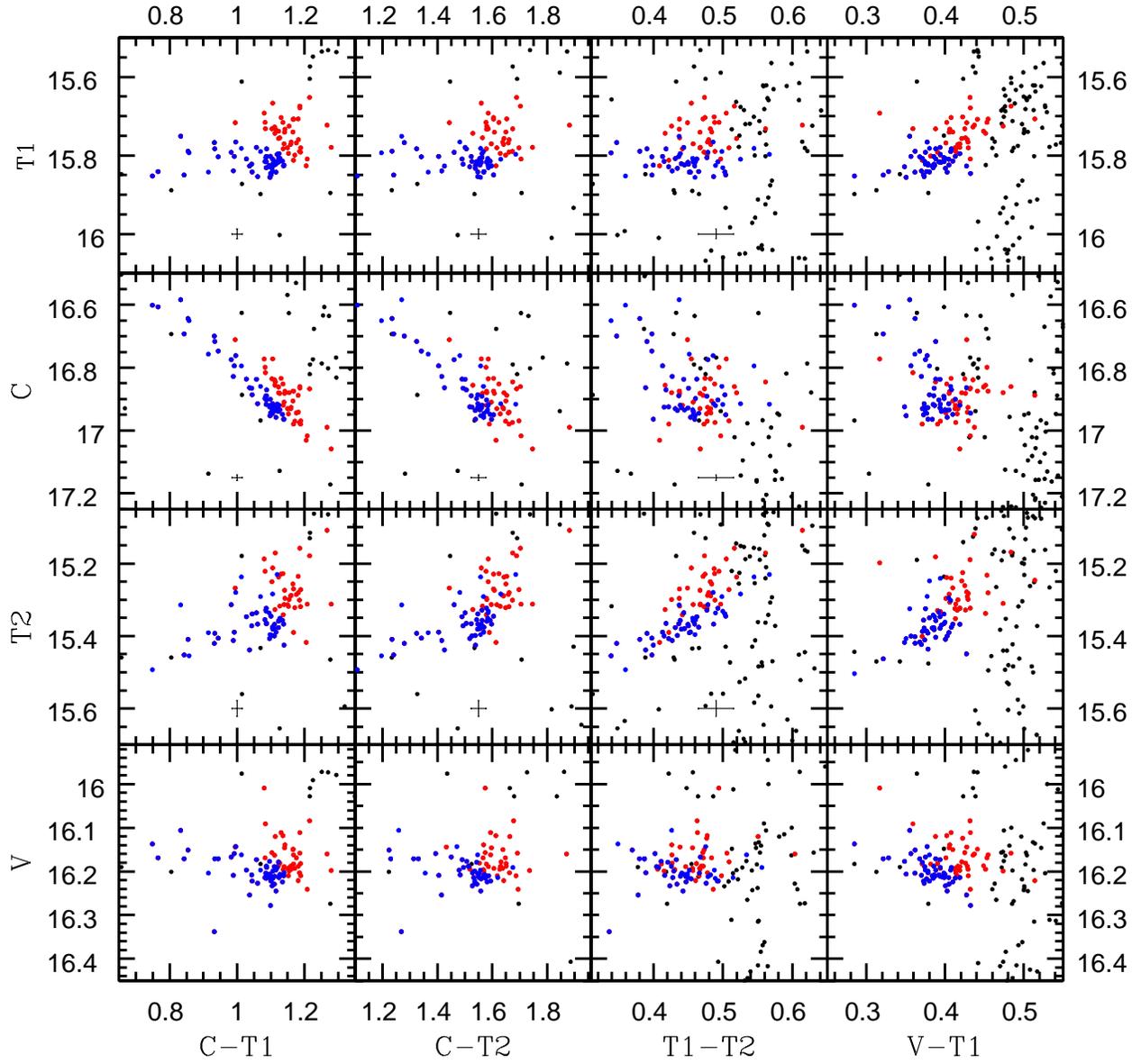}
\end{center}
\vspace{-5cm}
\caption{The double red horizontal branch in all of the filter combinations.  The red and blue
colors are defined by the two sequences observed in C-T1 versus T1 and then applied to
all of the other CMDs.}
\end{figure}
\clearpage

\begin{figure}[htp]
\begin{center}
\includegraphics[scale=0.8]{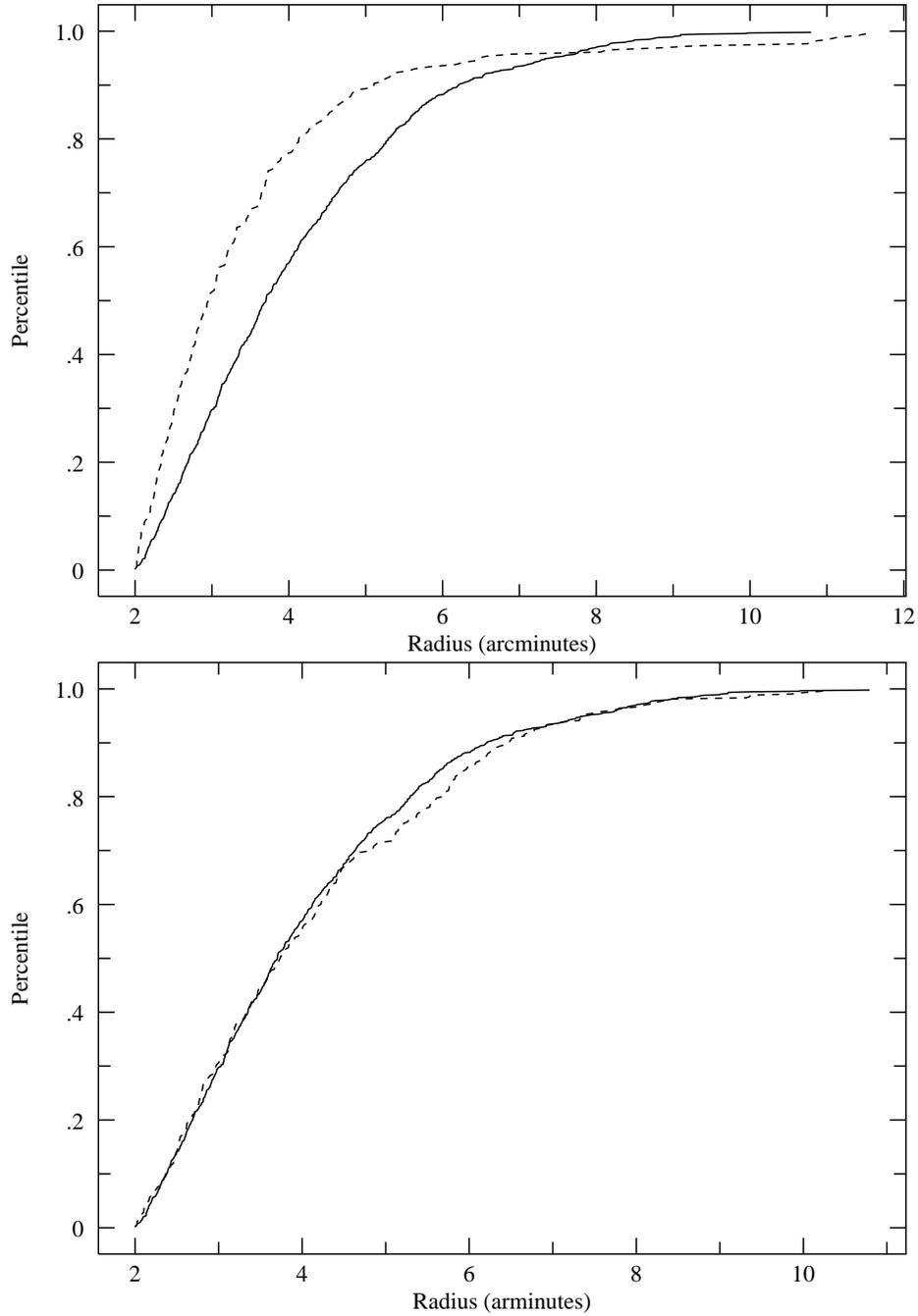}
\end{center}
\vspace{-2.5cm}
\caption{The upper panel illustrates our KS test for the red MS (dashed line) and the blue MS (solid line) populations.  
Due to the effects of crowding in the inner core, we only examine stars greater than 2 arcminutes from the
core.  To limit the effects of the population overlap, we have only selected stars in the MS color distribution
(see left panel of Figure 8) from -.035 to 0.0 as the blue MS and from 0.085 to 0.030 as the red MS.  The
higher concentration of the red MS is shown clearly. The lower panel compares the same blue MS population (solid line)
to a redder sample (dashed line) that is still dominated by the blue MS, with a color distribution from 0.02 to 0.04.
There is no significant difference between the radial distributions in this case.}
\end{figure}
\clearpage

\begin{appendix}
\section{Analysis of Main Sequence Errors \& Artificial Stars}

Section 4.1 provides strong evidence that we have a distinct second redder MS population.
While we initially limited our MS sample to stars with relatively low C-T1 color error of $<$0.05,
we should still test if photometric errors could be playing a role by falsely creating
this red sequence.  For the two populations we have compared their color errors to their color
distribution and the corresponding blue MS $\sigma$.  The blue MS has a median color error 
of 0.034.  Its color distribution gives a $\sigma$ of 0.028 with a median color residual 
of 0.020, which can be sufficiently explained the blue MS errors.  The red MS has a comparable 
median color error of 0.037, which demonstrates its stars have no significant increase
in errors.  Direct comparisons of their color residual to color error for individual red-population 
stars gives ratios that range from 0.94 to 7.42 with a median ratio of 2.26.  Similarly, the 
red star color residuals range from 1.34 to 8.81$\sigma$ with a median value of 2.67$\sigma$.  Therefore, 
based on both color errors and distribution analysis, a majority of these red population stars have significant 
color differences from the MS fiducial.  

As with all ground-based observations of GCs, the effects of crowding should be considered, even 
in our relatively uncrowded outer-annulus observations of the MS of NGC 1851.
Based on the individual PSFs from our images we have randomly placed 50 artificial MS stars (19.75$<$C$<$20.75;
19.0$<$T1$<$20.0) across the full field with added Poisson noise.  This is a small enough number of stars to
not further increase the crowding, and to increase our sample we repeated this 20 times in both C
and T1.  Our photometric methods have independently detected in C all 269 of the artificial stars that fell within 
our outer annulus from 4 to 6.5 arcminutes, and in T1 it detected all 241 of the artificial stars that fell within our
outer annulus.  For comparison, across the entire field our photometric methods have independently detected in C
99\% of the input stars and in T1 99.4\% of the input stars.  One of the stars that was not detected fell directly
on a bright field star, and all other artificial stars that were not detected fell in the core within 2.0 arcminutes
of the center where it is more challenging to resolve faint MS stars.

Direct comparisons of the input and output magnitudes for our artificial stars provides a valuable tool to analyze 
our error characteristics.  It should be noted that this is based only on single measurements of individual images; 
therefore, the magnitude of these errors will not be representative of our final errors based on multiple combined 
measurements.  In our outer annulus we find in C the residual distribution has a $\sigma$ of 0.025, and 5.2\% of 
our sample have an observed difference greater than 0.1 ($>4\sigma$) from the input.  Similarly, in T1 we find the 
residual distribution has a $\sigma$ of 0.018, and 3.7\% of our sample have an observed difference greater than 
0.072 ($>4\sigma$) from the input.  The larger effects seen in C are likely due to both the MS stars being fainter 
in C and the larger seeing in these observations, resulting in an increased probability of being affected by 
crowding.  It should be noted that nearly all (87\%) of these magnitude outliers are found to be brighter than 
their input magnitudes, consistent with them being very near another star that adds light to its measurement.
We take these low percentages to be our probability of being affected by crowding for the full sample of MS 
stars in this outer annulus.  However, these magnitude outliers also exhibit moderately larger photometric errors and 
will have larger magnitude dispersions across multiple images, leading them to have been likely cut based on errors 
in our MS analysis here.  Furthermore, if these crowded stars are not cut by our error requirements, the increased 
effect of crowding on C over T1 will lead crowded stars to be preferentially bluer, in stark contrast to our clearly 
redder C-T1 population.  Consistent with this, the cut applied to our initial sample (see Section 2) of stars 
within 2 arcminutes of the center, where crowding is far more serious, shows that the cut stars with large errors were 
predominantly bluer than the primary RGB and MS.  A further test for the potential effects of crowding is performed 
by directly marking the two populations in an image of the cluster.  We see that in this outer annulus the red MS 
population are not more likely to have near neighbors that could cause problems and they are primarily isolated 
stars.  Reassuringly, there also is no clear difference in their spatial distributions azimuthally.
\end{appendix}

\end{document}